\documentclass[iop]{emulateapj}
\usepackage{enumitem}

\begin{document}

\newcommand{\myemail}{jsohn@cfa.harvard.edu}
\newcommand{\kms}{\rm km~s^{-1}}
\newcommand{\kmsmpc}{\rm km~s^{-1}~Mpc^{-1}}
\newcommand{\ergcms}{\rm erg~cm^{-2}~s^{-1}}
\newcommand{\dn}{D_{n}4000}

\newcounter{mytempeqncnt}

\title{A Spectroscopic Census of X-ray Systems in the COSMOS Field}

\author{Jubee Sohn$^{1}$,
             Margaret J. Geller$^{1}$, H. Jabran Zahid$^{1}$}

\affil{$^{1}$ Smithsonian Astrophysical Observatory, 60 Garden Street, Cambridge, MA 02138, USA}

\begin{abstract}
We investigate spectroscopic properties of galaxy systems identified based on deep X-ray observations in the COSMOS field.
The COSMOS X-ray system catalog we use \citep{George11} includes 180 X-ray systems to a limiting flux of $1.0 \times 10^{-15} \ergcms$,
 an order of magnitude deeper than future e-ROSITA survey.
We identify spectroscopic members of these X-ray systems
 based on the spectroscopic catalog constructed by compiling various spectroscopic surveys including 277 new measurements;
 137 X-ray systems are spectroscopically identified groups with more than three spectroscopic members.
We identify 1843 spectroscopic redshifts of member candidates in these X-ray systems.
The X-ray luminosity ($L_{X}$) - velocity dispersion ($\sigma_{v}$) scaling relation of the COSMOS X-ray systems is consistent with 
 that of massive X-ray clusters. 
One of the distinctive features of the COSMOS survey is that it covers the X-ray luminosity range
 where poor groups overlap the range for extended emission associated with individual quiescent galaxies. 
We assess the challenges posed by the complex morphology of the distribution of low X-ray luminosity systems, including groups and individual quiescent galaxies, 
 in the $L_{x} - \sigma_{v}$ plane.
\end{abstract}

\section{INTRODUCTION}

Detection of hot X-ray emitting gas from galaxy clusters
 was a groundbreaking discovery that provided
 a powerful tool for studying the formation and evolution of gravitationally bound galaxy systems
 (see review by \citet{Sarazin88}).
The X-ray observations trace extended thermal emission from the intergalactic medium in galaxy systems (e.g. \citealp{Kellogg72, Forman72}).
The extended X-ray emission is detected not only in massive galaxy clusters  
 but also in less massive galaxy groups \citep{Ponman96, Mahdavi00}. 
Hot X-ray emitting gas is also detected around individual galaxies \citep{Fabbiano89, Kim18}.
The X-ray luminosity and temperature of the hot gas
 is well-correlated with the mass of its parent dark matter halo
 (e.g. \citealp{Stanek06, Bohringer10}).

A simple theoretical calculation based on the hydrostatic equilibrium predicts
 the scaling relations between X-ray properties and the other physical properties of a galaxy system.
For example,
 the X-ray luminosity ($L_{X}$) and the X-ray temperature ($T_{X}$) of a system scale with
 the measured velocity dispersion \citep{Solinger72}.
Studying scaling relations probe the underlying cluster physics.
Extension of the scaling relations to a wider X-ray luminosity range
 allows for exploration of the nature of X-ray emission from lower mass systems. 
Extension to a broader redshift range probes the evolution of galaxy systems.

A number of systematic surveys identify galaxy systems in the X-ray 
 (e.g. \citealp{Ebeling98, Bohringer00, Bohringer17, Reiprich02, Burenin07, Pacaud16}).
Verification of these X-ray systems depends on optical imaging combined with spectroscopic observations.
For example, SDSS spectroscopy combined with {\it ROSAT} and {\it XMM} imaging \citep{Clerc16} yields a large catalog of galaxy clusters.
\citet{Sohn18b} used the deeper HectoMAP redshift survey \citep{Geller15} and {\it ROSAT} X-ray imaging to identify galaxy clusters.

The search for galaxy systems also extends to lower mass galaxy groups.
AEGIS (All-Wavelength Extended Groth Internation Survey, \citealp{Erfanianfar13}) describes
 a catalog of galaxy systems including very low X-ray luminosity systems with a few spectroscopically identified members.
The XXL survey, based on XMM data, is a more extensive systematic survey for galaxy systems including low X-ray luminosity groups
 \citep{Pacaud16, Adami18}.
Many XXL galaxy groups are confirmed based on extensive spectroscopic surveys.
In general, X-ray systems verified with the spectroscopic redshifts
 provide an important basis for studying the nature and evolution of galaxy systems.

COSMOS \citep{Scoville07} is a unique field where extensive multi-wavelength data 
 enables a study of X-ray systems with a large range of X-ray luminosity and redshift.
The COSMOS survey covers a $\sim 2$ deg$^{2}$ field.
The X-ray observations for the COSMOS field are very deep
 ($f_{X} = 3 \times 10^{-16}$ erg s$^{-1}$ cm$^{-2}$, \citealp{Gozaliasl19}),
 an order of magnitude deeper than the future e-ROSITA survey \citep{Merloni12}.
Based on these deep X-ray observations,
 \citet{Finoguenov07} first identify extended X-ray emission corresponding to  galaxy groups and clusters.
\citet{George11} and \citet{Gozaliasl19} extended the search for X-ray systems with even deeper X-ray data.
These studies also identify optical counterparts of the extended X-ray sources.
However, these analyses have relied primarily on photometric redshifts.

Dense spectroscopic surveys covering the field add an additional dimension to the COSMOS field.
zCOSMOS is the first of these systematic spectroscopic surveys \citep{Lilly07, Lilly09}.
Several large spectroscopic surveys include high redshift objects
 in the COSMOS field (e.g. \citealp{Silverman15, Hasinger18}). 
Remarkably these deep surveys
 include only very sparse coverage for redshifts $z \lesssim 0.7$.
To remedy this situation, hCOSMOS \citep{Damjanov18} is a shallower, 
 but complete ($> 90\%$) redshift survey to $r = 20.6$
 covering the central $\sim 1$ deg$^{2}$ of the COSMOS field.
Taken together, these redshift surveys provide an opportunity
 for a spectroscopic census of the galaxy system candidates.

Here, we examine extended X-ray sources as candidate galaxy systems in the COSMOS field.
We combine the COSMOS galaxy catalog \citep{George11} and
 all available spectroscopic redshifts for the COSMOS field.
We identify spectroscopic members associated with X-ray systems.
We also identify the brightest galaxy in the system and
 we examine the scaling relation between the X-ray luminosity and the velocity dispersion of the systems. 
The COSMOS X-ray data are so deep that 
 they begin to probe the luminosity range where individual quiescent galaxies and 
 poor systems overlap in X-ray luminosity. 
We highlight some of the complexity introduced by this overlap.

We describe the observations in Section \ref{data}.
We demonstrate the spectroscopic survey completeness in Section \ref{completeness}.
In Section \ref{catalog}, we explore the spectroscopic catalog of the COSMOS X-ray systems
 including spectroscopic membership identification. 
We construct the X-ray scaling ($L_{X} - \sigma_{v}$) relation in Section \ref{relation}. 
In Section \ref{discussion} we consider some of the limitations of the analysis and 
 we highlight the complex morphology of the distribution of extended X-ray sources in the $L_{X} - \sigma_{v}$ plane. 
We summarize in Section \ref{conclusion}.
In the appendix, we supplement the X-ray ID with a brief overview of corresponding photometrically identified clusters.
We use the standard $\Lambda$CDM cosmology with $H_{0} = 70~\kmsmpc$, $\Omega_{m} = 0.3$, and $\Omega_{\Lambda} = 0.7$
 throughout.

\section{THE DATA}\label{data}

COSMOS is a deep multi-wavelength survey \citep{Scoville07}
 covering a 2 deg$^{2}$ field at (R.A., Decl. ) = (150.1192, +2.2058).
The field has been observed over a wide range of wavelengths from X-ray to radio
 with major observing facilities.
Here, we use optical photometry and spectroscopy to investigate
 galaxies in candidate COSMOS X-ray systems.
We describe the galaxy catalog and photometry in Section \ref{phot},
 the spectroscopic data in Section \ref{spec},
 and the COSMOS spectroscopic sample that combines photometric and spectroscopic data in Section \ref{sample}.
In Section \ref{xcat}, we describe the X-ray system catalogs we use.

\subsection{Galaxy Catalog}\label{phot}

We use the COSMOS Galaxy and X-ray Group Membership Catalog\footnote{https://irsa.ipac.caltech.edu/cgi-bin/Gator/nph-dd?catalog=cosmos\_xgal}
 \citep{George11} as the basis of our study.
This catalog includes 115,844 galaxies with $m_{\rm F814} < 24.2$ (MAG\_AUTO)
 from the COSMOS Advanced Camera for Survey (ACS) catalog \citep{Leauthaud07}.
The catalog also lists photometric redshifts \citep{Ilbert09} and
 membership probabilities for galaxies in the candidate X-ray systems (Section \ref{xcat}, \citealp{George11}).
Hereafter, we refer to this catalog as the COSMOS galaxy catalog.

We match the COSMOS galaxy catalog to the UltraVISTA photometric catalog \citep{Muzzin13}
 to obtain optical photometry of the galaxies.
The UltraVISTA photometric catalog provides point source function (PSF) matching photometry over 30 photometric bands.
We use a $0.5\arcsec$ search radius for matching.
We obtain {\it Subaru} $g^{+}r^{+}i^{+}z^{+}y^{+}$ photometry of 110,375 galaxies from the UltraVISTA catalog.
There are a number of objects without UltraVISTA photometry
 due to the small field of view of the UltraVISTA survey relative to the COSMOS galaxy catalog.

We supplement galaxy magnitudes using the COSMOS photometry catalog
 \footnote{https://irsa.ipac.caltech.edu/cgi-bin/Gator/nph-dd?catalog=cosmos\_phot} \citep{Capak07}, 
{\bf because many objects in the COSMOS galaxy catalog are located outside the UltraVISTA coverage. }
This catalog also offers {\it Subaru} $g^{+}r^{+}i^{+}z^{+}y^{+}$ photometry of the galaxies in the COSMOS field.
However, the magnitudes from the COSMOS photometry catalog are not identical to the UltraVISTA photometry.
Therefore, we need to transform the magnitudes from the COSMOS photometry catalog ($m_{\rm C07}$) into the UltraVISTA magnitudes ($m_{\rm UVIS}$).

Based on the 24591 bright galaxies ($16 \leq r^{+} \leq 23$) with both COSMOS and UltraVISTA photometry,
 we derive empirical transformations from $m_{\rm C07}$ into $m_{\rm UVIS}$.
For example, the $r-$band magnitude transformation is:
\begin{equation}
r^{+}_{\rm UVIS} = (1.11 \pm 0.01) \times r^{+}_{C07} - (2.93 \pm 0.21) .
\end{equation}
Because we use the photometry only
 for estimating the spectroscopic survey completeness and
 for identifying the brightest cluster galaxies, the empirical transformation is sufficient.

\subsection{Spectroscopy}\label{spec}

We use spectroscopy to determine the membership of the COSMOS X-ray systems.
We first compile spectroscopic redshifts from several surveys covering the COSMOS field (Section \ref{surveys}).
In addition, we carried out spectroscopic observations to obtain additional redshifts in the field (Section \ref{obs}).

\subsubsection{Previous COSMOS Redshift Surveys}\label{surveys}
zCOSMOS \citep{Lilly07, Lilly09} is the largest spectroscopic survey covering the COSMOS field
 using VIMOS mounted on VLT/UT 8m telescope.
The zCOSMOS DR3 catalog we use lists 20689 redshifts of galaxies with $I_{AB} < 22.5$
 with an average accuracy of $\sim 110~\kms$.
zCOSMOS provided a confidence class ($z_{conf}$) for the redshift measurements.
We include zCOSMOS redshifts with high confidence class ($z_{conf} \geq 3$).
Using a $0.5\arcsec$ matching tolerance,
 we obtain 13935 redshifts for galaxies in the COSMOS galaxy catalog.

hCOSMOS \citep{Damjanov18} is a dense spectroscopic survey
 of a magnitude-limited sample with $r \leq 21.3$ in the COSMOS field.
The hCOSMOS survey was done with the multi-fiber fed spectrograph Hectospec \citep{Fabricant05} mounted on the MMT 6.5 m telescope.
The hCOSMOS survey includes 4362 redshifts within the central 1 deg$^{2}$ field;
 1701 redshifts are new.
\citet{Damjanov18} showed that
 the average difference between hCOSMOS and zCOSMOS redshift measurement is $\sim 17 \pm 2~\kms$,
 smaller than the typical uncertainty in each redshift measurement ($\sim 35~\kms$).
Thus, we use redshifts from zCOSMOS and hCOSMOS without correction.
We compile 4399 redshifts from the hCOSMOS catalog.

\citet{Hasinger18} provide a catalog of 10718 objects in the COSMOS field
 including 6617 objects with high-quality spectra
 observed with the DEep Imaging Multi-Object Spectrograph (DEIMOS) on the Keck II telescope.
Hereafter, we refer to this survey as dCOSMOS.
Similar to zCOSMOS, dCOSMOS provides a confidence class for the redshift measurements.
Following the zCOSMOS matching process, we obtain 2672 dCOSMOS redshifts with a high confidence flag ($z_{conf} \geq 3$).
There are 1185 dCOSMOS objects with zCOSMOS redshift measurements.
We compute the mean difference in the redshift measurement using these objects.
The typical offset between dCOSMOS and zCOSMOS redshifts is $\sim 12~\kms$
 and the typical uncertainty of the dCOSMOS redshifts is $\sim 60~\kms$ (inferred from other DEIMOS observations).

Besides these three large surveys,
 there are several additional spectroscopic surveys covering the COSMOS field.
We also obtain the redshifts from these surveys. 
Compiling the redshift measurements from these previous surveys is not straightforward.
These surveys obtained spectra using different instruments
 that yield spectra with varying spectral resolution and wavelength coverage.
Therefore, there may be some systematic differences between these redshift measurements
 and those from zCOSMOS, hCOSMOS and dCOSMOS.
Furthermore, the redshift measurements from these surveys may have large uncertainties
 which can compromise membership determination.
These surveys often omit individual uncertainties from their catalogs.

\begin{deluxetable*}{lcccc}
\tablecolumns{5}
\tabletypesize{\footnotesize}
\setlength{\tabcolsep}{0.05in}
\tablecaption{Spectroscopic Surveys of the COSMOS field}
\tablehead{
\colhead{Reference} & \colhead{$N_{z}$\tablenotemark{a}} & \colhead{$N_{z} (P_{mem} > 0)$\tablenotemark{b}} &
\colhead{$N_{overlap}$ (zCOS)\tablenotemark{c}} & \colhead{mean $\Delta cz_{\rm zCOS}~(\kms)$\tablenotemark{d}}}
\startdata
\citet{Prescott06}		&	247	&   13 &   85 & $  62\pm    8$ \\
\citet{Trump09}		&	387	&   27 &   90 & $ -54 \pm   16$ \\
\citet{Balogh14}		&	534	&  130 &   18 & $ -35 \pm   28$ \\
\citet{Comparat15}	&	1383	&   36 &  154 & $  27 \pm    10$\\
\citet{Kartaltepe15}	&	116	&    4 &   43 & $ -51 \pm   23$\\
\citet{Silverman15}	&	270	&    0 &    3 & $-103 \pm    4$\\
\citet{Masters17}		&	554	&   22 &   12 & $  14 \pm   41$ \\
\citet{Straatman18}	&	1699	&	141 &  554 & $  13 \pm    4$
\enddata
\label{speccat}
\tablenotetext{a}{Number of redshifts. }
\tablenotetext{b}{Number of redshifts for  member candidates of the COSMOS X-ray systems with $P_{mem} > 0$. }
\tablenotetext{c}{Number of objects with a redshift from the zCOSMOS survey.}
\tablenotetext{d}{Mean difference between the redshift measurement from the reference and zCOSMOS normalized by the zCOSMOS redshift.
 The mean difference was estimated after $3\sigma$ clipping of outliers. }
\tablenotetext{e}{Same as (c), but for the hCOSMOS survey.}
\tablenotetext{f}{Same as (d), but for hCOSMOS redshift measurement.}
\end{deluxetable*}

Table \ref{speccat} summarizes the previous spectroscopic surveys we compile and 
 lists the number of redshifts for galaxies and member candidates of the X-ray groups/clusters.
We also investigate the number of objects in each of these previous spectroscopic surveys that also have a zCOSMOS redshift.
Based on these repeat measurements,
 we calculate the systematic differences between the redshift measurements
 from zCOSMOS and the other surveys ($\Delta cz / (1 + z_{\rm zCOSMOS})$).
We briefly review these previous spectroscopic surveys.

\begin{itemize}[leftmargin=*]
\item[] {\bf \citet{Prescott06}}:
This survey measured redshifts of galaxies and quasars in the COSMOS field using MMT/Hectospec.
We obtained 247 redshifts from this survey for objects in the COSMOS galaxy catalog.
The typical uncertainty of the redshift measurements is $\sim 63~\kms$.
zCOSMOS also measured redshifts for 85 objects from this survey.
We compute the difference between the redshift (radial velocity) measurements ($\Delta cz / (1 + z_{\rm zCOSMOS})$)
 from this survey and zCOSMOS.
The mean difference estimated after $3\sigma$ clipping is $62 \pm 8~\kms$.
Because the redshift difference is smaller than the uncertainty in the zCOSMOS redshift measurement,
 we do not apply a systematic correction.

\item[] {\bf \citet{Trump09}: }
This survey provided optical spectroscopy for X-ray point-like sources identified from the {\it XMM-Newton} data.
The majority ($\sim 2/3$) of spectra were obtained with the Inamori Magellan Areal Camera and Spectrograph (IMACS) on the Magellan (Baade) telescope,
 and the rest of the spectra were obtained with MMT/Hectospec or SDSS.
\citet{Trump09} provided a confidence flag ($z_{conf}$) for the redshift measurements.
We take 387 redshifts with $z_{conf} = 3$ or 4,
 where $z_{conf} = 3$ indicates that the redshift was measured based on one strong and one weak features,
 and $z_{conf} = 4$ indicates that the redshift is considered unambiguous.
Based on the 90 objects with repeat redshift measurements,
 we computed a mean difference of $-54 \pm 16~\kms$ compared with the zCOSMOS redshifts.

\item[] {\bf The GEEC2 spectroscopic survey} \citep{Balogh14}:
The Galaxy Environment Evolution Collaboration 2 (GEEC2) spectroscopic survey presented
 a catalog based on the Gemini-South GMOS spectroscopy of 11 galaxy groups at $0.8 < z < 1$ in the COSMOS field.
This catalog lists 603 unique redshifts with high confidence and with a typical uncertainty of $\sim 100~\kms$.
\citet{Balogh14} claim that there is an $\sim 180~\kms$ unexplained systematic offset compared to zCOSMOS redshifts.
However, we only observe a much smaller offset ($-35 \pm 28~\kms$) based on 18 repeat measurements
 with high confidence ($z_{conf} > 3$),
 smaller than typical uncertainties in both GEEC2 and zCOSMOS redshift measurements.
Therefore, we compile GEEC2 redshifts without correction.

\item[] {\bf \citet{Comparat15}}:
This survey used FOcal Reducer and the low dispersion Spectrograph (FORS2) for  the Very Large Telescope (VLT)
 to obtain the redshifts and emission-line fluxes of galaxies at $0.1 < z < 1.65$.
\citet{Comparat15} investigated the [O II] luminosity function using this spectroscopic sample.
The redshift catalog includes a confidence flag for the redshift measurements similar to the zCOSMOS confidence flag.
We compile 1383 redshifts with $z_{conf} = 3$ or 4.
The mean redshift difference from zCOSMOS redshifts measured from 154 duplicated objects
 is $27 \pm 10~\kms$.

\item[] {\bf \citet{Kartaltepe15}}:
The FMOS-COSMOS survey published redshifts measured from the FMOS near-infrared spectrograph on the {\it Subaru} telescope.
From this survey, we compile 116 redshifts within the range $0.5 < z < 1.7$.
There is a small overlap (43 objects) with the zCOSMOS sample:
 the mean redshift difference is $-51 \pm 23~\kms$.

\item[] {\bf \citet{Silverman15}}:
We obtain redshifts from \citet{Silverman15}, who also published redshifts from the FMOS-COSMOS survey over the redshift range $0.7 < z < 2.5$.
There are 270 redshifts from this catalog, but none are X-ray system member candidates (Section \ref{xcat}).
The overlap of this sample with zCOSMOS is negligible.
The typical velocity resolution from the FMOS spectroscopy is $\sim 115~\kms$ \citep{Silverman15}.

\item[] {\bf \citet{Masters17}}:
The Complete Calibration of the Color–Redshift Relation (C3R2) Survey is a deep redshift survey
 that aims to calibrate the color-redshift relation of galaxies to the {\it Euclid} depth.
\citet{Masters17} presented Data Release (DR) 1 including redshifts of COSMOS galaxies measured from Keck spectroscopy.
We obtain 554 redshifts from DR1; only 12 of them overlap with zCOSMOS measurements.

\item[] {\bf \citet{Straatman18}}:
The Large Early Galaxy Astrophysics Census (LEGA-C) Data Release 2 provided redshifts from VLT/VIMOS observations.
LEGA-C includes 1699 redshifts for objects in the COSMOS galaxy catalog;
 554 objects also have redshifts from zCOSMOS.
The LEGA-C redshifts show a small systematic offset compared to zCOSMOS redshift of $13 \pm 4~\kms$.
\end{itemize}

\subsubsection{Observations}\label{obs}
We carried out new Hectospec observations for the COSMOS field in 2018 December.
Following the hCOSMOS survey,
 we used the 270 line mm$^{-1}$ Hectospec grating which yields a 6.2\AA~ spectral resolution over the wavelength range 3800 - 9100\AA.
The exposure time for the Hectospec observations is an hour.

The Hectospec data were reduced with the IDL HSRED v2.0 package.
We measure redshifts with RVSAO \citep{Kurtz98} that cross-correlates observed spectra with a set of redshift templates.
We visually inspect the cross-correlation results and classified them into three groups:
 ‘Q’ for high-quality fits, ‘?’ for ambiguous cases, and ‘X’ for poor fits.
We use 277 redshifts derived from high-quality fits with a typical uncertainty of $\sim 36~\kms$;
 52 of them are X-ray system member candidates.
Table \ref{redshift} lists all of the 277 redshifts we obtained from the new Hectospec observations.

\begin{deluxetable}{cccc}
\tablecolumns{4}
\tabletypesize{\footnotesize}
\setlength{\tabcolsep}{0.1in}
\tablecaption{New Spectroscopic Redshifts from Hectospec}
\tablehead{\colhead{R.A.} & \colhead{Decl.} & \colhead{z} & \colhead{z error}}
\startdata
149.556005 &   2.362138 &  0.37433 &  0.00019 \\
149.914532 &   2.450852 &  0.44304 &  0.00030 \\
150.640850 &   2.466479 &  0.61610 &  0.00006 \\
150.683432 &   2.224703 &  0.41087 &  0.00013 \\
149.538039 &   2.073853 &  0.20916 &  0.00006
\enddata
\label{redshift}
\tablecomments{The entire table is available in machine-readable form in the online journal. }
\end{deluxetable}

\subsection{The COSMOS Spectroscopic Sample}\label{sample}

Based on the COSMOS galaxy catalog,
 we compile spectroscopic redshift measurements from a wide variety of surveys in the literature.
In addition, we obtain new redshifts from MMT/Hectospec observations.
As a result, we have 22667 redshifts for objects in the COSMOS galaxy catalog regardless of the brightness of the galaxies.
This COSMOS galaxy catalog includes 7600 galaxies with $P_{mem} > 0$,
 where $P_{mem}$ indicates the membership probability
 (\citealp{George11}, see details in Section \ref{xcat}).
Our compilation includes 1611 ($\sim 21\%$) spectroscopic redshifts for X-ray system member candidates.

We compare the spectroscopic redshifts with the photometric redshifts from the COSMOS galaxy catalog \citep{George11}.
These photometric redshifts were originally derived by \citet{Ilbert09}
 who compared spectral energy distributions (SEDs) from over 30 bands of Ultraviolet, optical and infrared data with a set of galaxy templates with stellar population synthesis models.
The photometric redshifts from \citet{George11} were updated with H-band data and with improved template fitting techniques.
The uncertainty in the photometric redshift measurement is $\Delta z \sim 0.008$.

\begin{figure}
\centering
\includegraphics[scale=0.5]{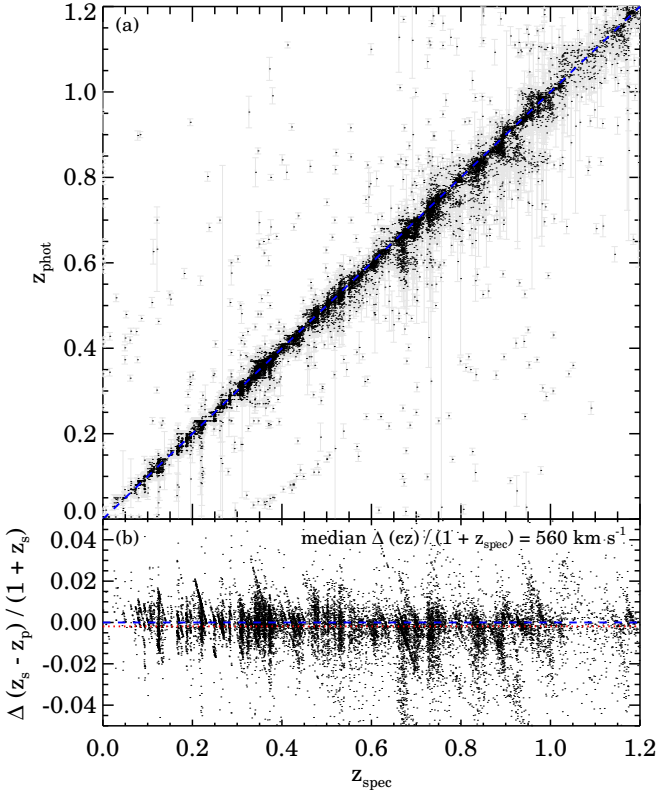}
\caption{(a) Comparison between photometric redshifts \citep{George11} and spectroscopic redshifts of galaxies in the COSMOS field.
The dashed-line shows the one-to-one relation.
(b) The difference between photometric- and spectroscopic redshifts as a function of spectroscopic redshift.
The red dotted line shows the median redshift difference and blue dashed line shows zero offset.
The median redshift difference is $-560 \pm 120~\kms$, larger than the typical velocity dispersion of an X-ray system. }
\label{zphot}
\end{figure}

Figure \ref{zphot}(a) compares photometric and spectroscopic redshifts
 for the COSMOS objects.
In Figure \ref{zphot}(b), we display the difference between the photometric and spectroscopic redshifts
 normalized by the spectroscopic redshift.
The median difference (the red dotted line) over this redshift range is $-560 \pm 120~\kms$
 smaller than the typical uncertainty in the photometric redshift;
 it is interesting that there is systematic offset between photometric and spectroscopic redshifts.
This large difference suggests that 
 membership identification for group and clusters based on photometric redshifts may be incorrect.

We also estimate the stellar mass of the galaxies with spectroscopic redshifts
 based on the technique used in \citet{Damjanov18}.
We use the Le Phare fitting code \citep{Arnouts99, Ilbert06}
 to derive a mass-to-light ratio by comparing observed magnitudes with a spectral energy distribution model.
We use optical $ug^{+}r^{+}i^{+}z^{+}y^{+}$ band magnitudes of individual galaxies.
For comparison,
 we generate a set of synthetic SED model based on the stellar population synthesis model from \citet{BC03} and a \citet{Chabrier03} IMF.
We assume exponentially declining star forming histories with  e-folding time scales  of $\tau = 0.1, 0.3, 1, 2, 3, 5, 10, 15$ and 30 Gyr
 and three metallicities ($Z = 0.004, 0.008$ and 0.02).
We also take into account the variation of the stellar population age in the range 0.01 to 13 Gyr.
We account for foreground extinction using the \citet{Calzetti00} extinction law with an $E(B-V)$ range of 0.0 to 0.6.
We obtain the median value of the stellar mass from the probability distribution function for the stellar mass.

\subsection{COSMOS X-ray Systems and Their Members}\label{xcat}

We use the COSMOS X-ray group catalog from \citet{George11}.
This catalog is the second version of the catalog of X-ray systems in the COSMOS field
 updated from \citet{Finoguenov07} and was followed by \citet{Gozaliasl19}.
We use the catalog from \citet{George11} because they included a catalog of X-ray system candidate member galaxies
 along with the catalog of X-ray systems.
This member candidate catalog is critical for determining the spectroscopic membership
 based on spectroscopic survey data.

The first catalog of X-ray clusters in the COSMOS field is from \citet{Finoguenov07}.
They use the first 36 {\it XMM-Newton} pointings covering a 2.1 deg$^{2}$ area of the COSMOS field.
They identify 72 X-ray cluster candidates with extended X-ray emission associated with concentrations of galaxies
 in photometric redshift space.
The X-ray survey reaches a flux limit of $3 \times 10^{-15} \ergcms$
 within the $0.5-2.0$ keV band.
They provide the physical properties of the X-ray systems including X-ray luminosity, temperature and
 characteristic mass ($M_{500}$) based on X-ray scaling relations.

\citet{George11} extended the search for X-ray systems
 based on 54 {\it XMM-Newton} pointings and additional {\it Chandra} observations.
The X-ray flux limit of this survey is $\sim~ 1 \times 10^{-15} \ergcms$.
The deeper {\it XMM} coverage enables identification of fainter X-ray systems;
 \citet{George11} identify 211 extended X-ray sources over 1.64 deg$^{2}$;
 \citet{Finoguenov07} detect $\sim 150$ extended sources.
To identify the optical counterparts of the extended X-ray sources,
 \citet{George11} apply a red-sequence finder to galaxies within a projected distance of 0.5 Mpc
 from the X-ray center.
The X-ray system catalog from \citet{George11} includes
 165 X-ray groups and clusters with optical counterparts and
 18 extended X-ray sources with ambiguous optical counterparts.

The X-ray group membership catalog is a primary asset of \citet{George11}.
They employ a Bayesian approach to assign membership for galaxies
 associated with extended X-ray emission.
The selection algorithm determines the membership probability ($P_{mem}$)
 based on the position, the photometric redshift, and the stellar mass of the galaxy.
The COSMOS galaxy catalog includes 7600 galaxies with $P_{mem} > 0$ (4623 galaxies with $P_{mem} > 0.5$)
 regardless of the magnitude.

\citet{Gozaliasl19} is the most recent update for the COSMOS X-ray system survey.
They use even deeper{\it XMM-Newton} and {\it Chandra} observations and
 reach an X-ray flux limit of $3 \times 10^{-16} \ergcms$,
 an order of magnitude deeper than \citet{George11}.
They identify 247 X-ray groups in the redshift range $0.08 < z < 1.3$.
The major update from the previous catalogs is identification of higher redshift systems.
For systems at $z < 1.0$,
 the X-ray luminosities of the systems included in both \citet{George11} and \citet{Gozaliasl19} are essentially identical.
We use the properties of X-ray systems from \citet{George11}
 because we are limited by the spectroscopy to $z < 1$ and because \citet{Gozaliasl19}
do not provide a candidate member catalog.

The final catalog we use includes 180 X-ray systems at $z < 1$
 among 183 X-ray systems listed in \citet{George11}.
We exclude three systems because they have no member candidates ($P_{mem} > 0$) in the COSMOS galaxy catalog.
\citet{George11} offer a quality flag ($Q_{X}$) for the X-ray systems;
 1 indicates a well defined X-ray system, 2 indicates a well detected X-ray system with an ambiguous X-ray center, and
 3 indicates suspicious X-ray systems that require checking with spectroscopic data.
The three systems we exclude are $Q_{X} = 2$ systems.
There are 62 systems with $Q_{X} = 1$, 100 systems with $Q_{X} = 2$, and 18 systems with $Q_{X} = 3$. We begin by examining all of the systems with some members regardless of the quality flag.

\begin{figure}
\centering
\includegraphics[scale=0.5]{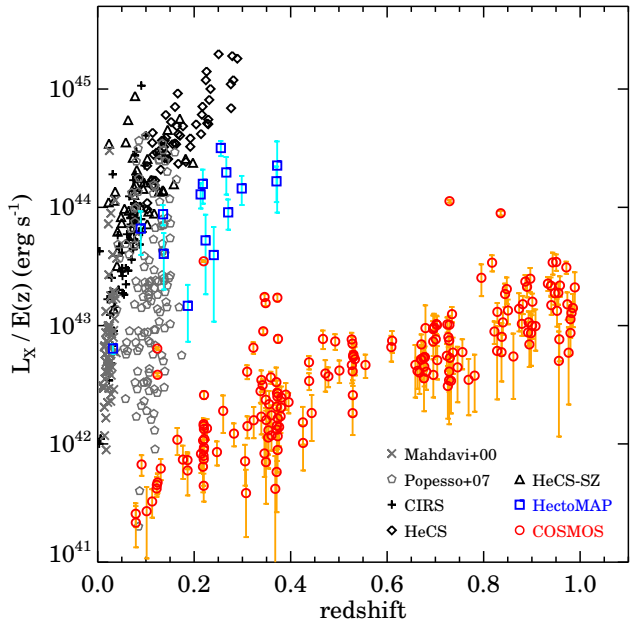}
\caption{X-ray luminosity of COSMOS X-ray systems (red circles) as a function of redshift.
Gray crosses and pentagons show X-ray samples from \citet{Mahdavi00} and \citet{Popesso07}, respectively.
Black pluses, diamonds, and triangles show the CIRS \citep{Rines03}, HeCS \citep{Rines13}, and HeCS-SZ \citep{Rines16} samples, respectively.
Blue circles show X-ray clusters in HectoMAP \citep{Sohn18b}. }
\label{lxz}
\end{figure}

Figure \ref{lxz} shows the X-ray luminosity as a function of redshift for all of the candidate COSMOS X-ray systems.
For comparison,
 we plot X-ray clusters with spectroscopy from the literature \citep{Mahdavi00, Popesso07, Rines03, Rines13, Rines16, Sohn18b}.
The COSMOS X-ray systems cover a wider redshift range than previous samples.
More importantly, the COSMOS systems include very low luminosity X-ray systems at a given redshift
 as a result of the very deep X-ray data.
In particular, at $z < 0.4$,
 the COSMOS X-ray sample offers a unique chance to study very low luminosity systems with $\log L_{X} < 42.5$.

\section{SPECTROSCOPIC SURVEY COMPLETENESS}\label{completeness}

\begin{figure}
\centering
\includegraphics[scale=0.5]{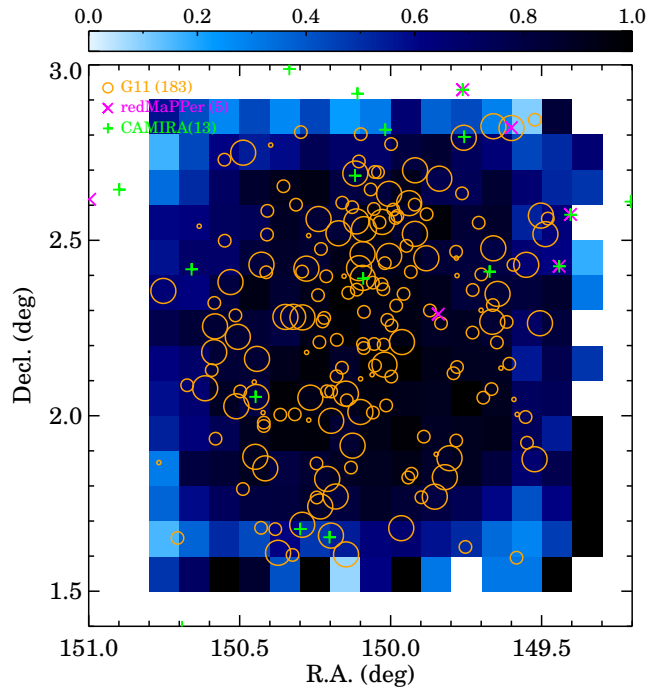}
\caption{Spatial distribution of galaxy cluster candidates in the COSMOS field.
The background map shows the spectroscopic completeness.
Circles display candidate X-ray systems with good X-ray flags.
In addition, the magenta crosses show redMaPPer clusters \citep{Rykoff16} and the green pluses display CAMIRA clusters \citep{Oguri14}.  }
\label{spatial}
\end{figure}

Figure \ref{spatial} shows the projected spatial distribution of the COSMOS X-ray systems.
The background density map displays the completeness of the combined spectroscopic surveys to $r = 21.3$.
Yellow circles show the location of the X-ray systems.
The largest symbols indicate a high X-ray confidence flag ($Q_{X} = 1$); the smallest symbols indicate $Q_{X} = 3$.
Most of the X-ray systems are located in the area where the spectroscopic survey is complete.
We also mark the location of galaxy cluster candidates based on red-sequence detection;
 magenta crosses indicate redMaPPer systems \citep{Rykoff16} and green pluses indicate CAMIRA clusters \citep{Oguri14}.
We describe these systems in Appendix \ref{photcl}.

\begin{figure}
\centering
\includegraphics[scale=0.5]{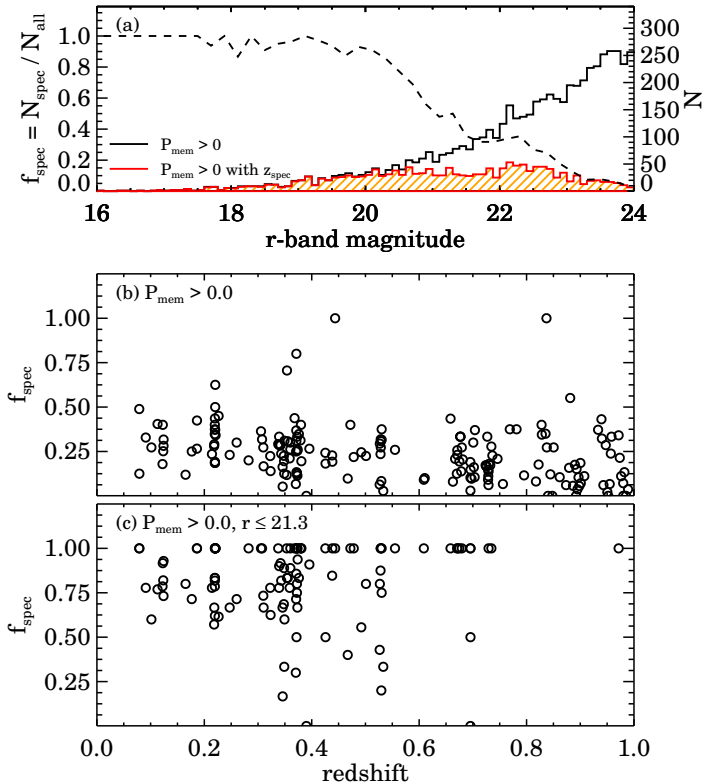}
\caption{(a)
Spectroscopic completeness (dashed line) of X-ray system member candidate ($P_{mem} > 0$) as a function of $r-$band magnitude.
In addition, histograms show the $r-$band magnitude distribution of all member candidates (open) and
 member candidates with spectroscopic redshifts (hatched).
(b) Spectroscopic completeness of individual X-ray systems:
 the ratio between $P_{mem} > 0$ galaxies with spectroscopic redshifts and all $P_{mem} > 0$ galaxies.
(c) Same as (b), but for the member candidates brighter than $r \leq 21.3$. }
\label{zcomp}
\end{figure}

Figure \ref{zcomp}(a) shows the spectroscopic survey completeness for the X-ray system member candidates with $P_{mem} > 0$.
We plot the $r-$band magnitude distribution of all member candidates along with the member candidates with spectroscopic redshifts.
To $r \sim 20.6$, the hCOSMOS survey limit, 91\% of the member candidates have spectroscopic redshifts.
The dashed line in Figure \ref{zcomp}(a) displays the spectroscopic completeness as a function of $r-$band magnitude;
 the completeness decreases rapidly for $r > 20.6$.

We compute the spectroscopic completeness for the individual X-ray systems.
In Figure \ref{zcomp}(b),
 we display the completeness for all galaxies with $P_{mem}$ in each X-ray system.
The overall completeness is typically low ($\sim 30\%$) due to missing redshifts for member candidates with $r > 21.3$
 (for every galaxy with $P_{mem} > 0$).
The spectroscopic survey for member candidates is much more complete for the brighter sample ($r \leq 21.3$, Figure \ref{zcomp} (c)).
For example, the spectroscopic survey is more than 75\% complete for 75 (42\%) of the COSMOS X-ray systems
 and more than 50\% complete for 95 (53\%) systems to $r = 21.3$.

\section{SPECTROSCOPIC CATALOGS OF COSMOS X-RAY SYSTEMS}\label{members}

We identify spectroscopic members of COSMOS X-ray systems based on the compilation of spectroscopic redshifts.
These spectroscopic members are the basis for studying the properties of the X-ray systems.
We describe the identification of spectroscopic members in Section \ref{identification}
 and show some example X-ray systems in Section \ref{example}.
We calibrate the photometric membership probability based on spectroscopy in Section \ref{calibration}.
We describe the final spectroscopic sample of COSMOS X-ray systems and
 the spectroscopic catalog of members in the COSMOS X-ray systems in Section \ref{catalog}.
In Section \ref{refined}, we define the refined sample based on X-ray flags and spectroscopic completeness  that we use for studying the $L_{X} - \sigma_{v}$ relation.
In Section \ref{bggidentification}, we identify the brightest group galaxies in the X-ray systems.

\subsection{Identification of Spectroscopic Members}\label{identification}

\begin{figure*}
\centering
\includegraphics[scale=0.6]{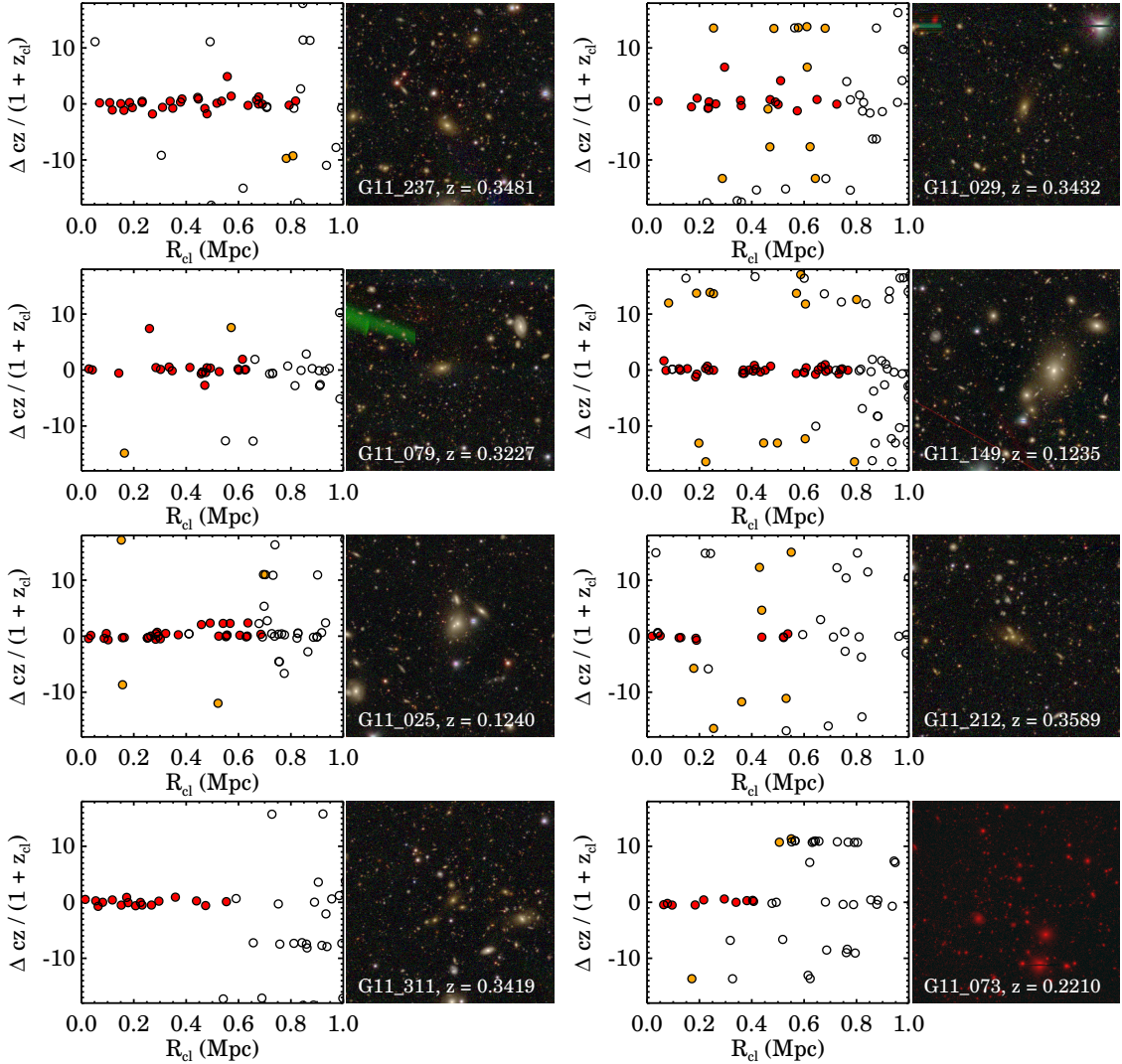}
\caption{
R-v diagrams of example COSMOS X-ray systems:
 the relative velocity difference with respect to the group mean redshift as a function of projected distance
 from the X-ray center.
The y-axis is in units of $1000~\kms$.
Red (orange) filled circles indicate galaxies with $P_{mem} > 0.5$ ($P_{mem} > 0$) and
 black open circles are galaxies with redshifts.
The right panels show the corresponding {\it Subaru}/Hyper Suprime Cam $gri$ images (500 kpc $\times$ 500 kpc). }
\label{rv}
\end{figure*}

We examine the distribution of X-ray system member candidates
 based on the classic R-v diagram, the rest-frame groupcentric velocity as a function of projected distance from the X-ray center.
Figure \ref{rv} shows  R-v diagrams of some well-sampled COSMOS X-ray systems:
 these systems are the brightest X-ray sources with more than 10 spectroscopic members at $z < 0.4$.
All of these example systems are used for studying $L_{X} - \sigma_{v}$ scaling relation (Section \ref{refined}).
The black open circles in Figure \ref{rv} show galaxies with spectroscopic redshifts around each X-ray system.
We also display the {\it Subaru}/Hyper Suprime Cam $gri$ image of each X-ray system
 within a 500 kpc $\times$ 500 kpc field of view.
The HSC images show that there are galaxy overdensities accompanying a dominant bright galaxy near the X-ray center.
Sometimes, the dominant bright galaxy is offset from the X-ray center.

In the R-v diagram,
 member candidates with $P_{mem} > 0.5$ (red circles) show a strong concentration around the center of each X-ray system.
In contrast, we note that a large fraction of the galaxies with $0 < P_{mem} < 0.5$ (orange circles) are line-of-sight interlopers.
They are barely distinguishable from cluster members based on photometric redshifts.

We identify spectroscopic members of the X-ray systems
 within $R_{proj} < R_{200, X}$.
$R_{200, X}$ is the characteristic radius of the X-ray system
 where the mean density is 200 times the critical density of the universe.
\citet{George11} estimated the $R_{200, X}$ based on the $L_{X} - M_{200}$ scaling relation
 calibrated based on the weak lensing mass \citep{Leauthaud10}.
We use the $R_{200, X}$ measurement because the membership probabilities from \citet{George11} are assigned to galaxies within this radius.

We also apply a generous $|\Delta c (z_{member} - z_{cl}) / (1 + z_{cl})| < 1000~\kms$ criterion to identify spectroscopic members.
Here the central redshift of the X-ray system ($z_{cl}$) is determined by iteration.
We first identify the spectroscopic member candidates based on a wider window:
 $R_{proj} < R_{200,X}$ and $|\Delta c (z_{member} - z_{cl}) / (1 + z_{cl})| < 3000~\kms$.
We take the median redshift of the member candidates as the spectroscopic redshift.
Then we finally identify the spectroscopic members with the narrower window of
 $R_{proj} < R_{200,X}$ and  $|\Delta c (z_{member} - z_{cl}) / (1 + z_{cl})| < 1000~\kms$.
The number of spectroscopic members changes little
 when we use a larger velocity difference limit of $1500$ or $2000~\kms$.

\begin{figure}
\centering
\includegraphics[scale=0.48]{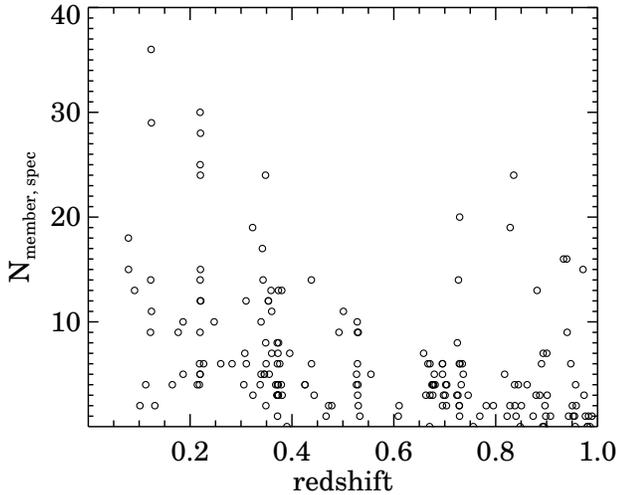}
\caption{Number of spectroscopic members of COSMOS X-ray systems as a function of redshift;
 76\% of the X-ray systems are spectroscopically identified groups with three or more members. }
\label{nmem}
\end{figure}

Figure \ref{nmem} shows the number of spectroscopic members of X-ray systems as a function of redshift.
We identify at least one spectroscopic members in 173 COSMOS X-ray systems.
There are 137 spectroscopically identified groups (76\%) with three or more spectroscopic members.
The median number of spectroscopic members is five,
 comparable with the XXL survey ($\sim 5$) combined with SDSS, GAMA and VLT spectroscopic survey \citep{Pacaud16}.
The systems where we do not identify any spectroscopic members are mostly (9/10) at high redshift ($z > 0.65$).

\subsection{Example COSMOS X-ray Systems}\label{example}

\begin{figure}
\centering
\includegraphics[scale=0.35]{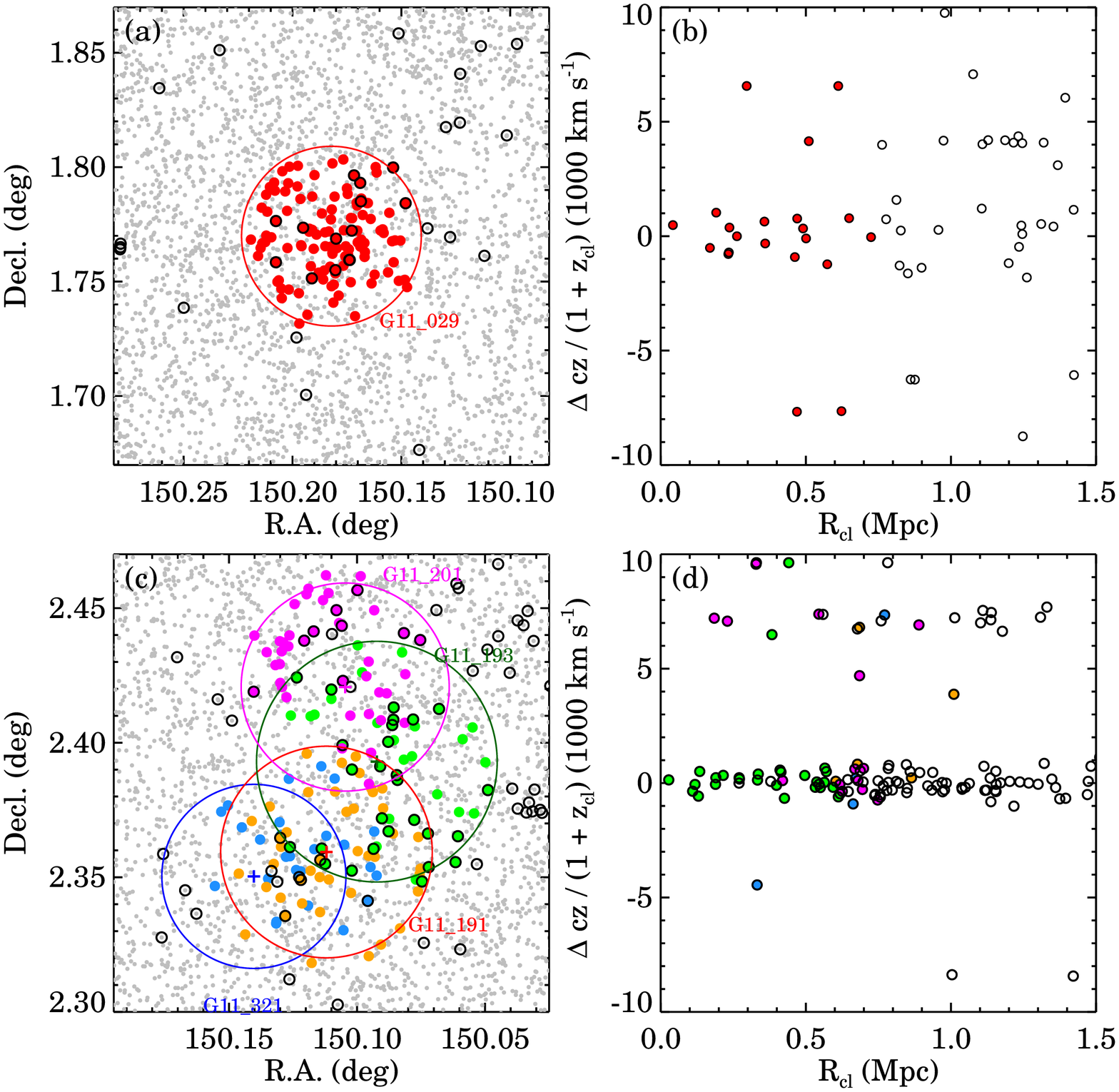}
\caption{
(a) Spatial distribution of galaxies around the well-defined X-ray system (ID: 029).
Gray dots are the galaxies in the field.
Filled circles are the photometrically identified members \citep{George11}
 and the open circles are 14 galaxies within $|\Delta c(z_{galaxy} - z_{system}) / (1. + z_{system})| < 1000~\kms$.
The large circle is $R_{200,X}$.
(b) R-v diagram for galaxies around the system shown in (a).
Filled circles are the member candidates with $P_{mem} > 0$, and open circles are galaxies with spectroscopic redsfhits.
(c) Same as (a), but for an overlapping set of four X-ray systems at $z \sim 0.22$.
Blue symbols are the members of X-ray system ID 321,
 red for ID 191, green for ID 193, and magenta for ID 201, respectively.
Open circles are the spectroscopically identified members with $|\Delta c(z_{galaxy} - z_{system}) / (1. + z_{system})| < 1000~\kms$.
(d) Same as (b), but for the overlapping four X-ray systems.
Filled circles are the member candidates in each X-ray system; color-coding is the same as (c).
The member candidates of the four apparent X-ray peaks are mixed together in the R-v diagram.}
\label{ovl}
\end{figure}

We explore two individual X-ray systems to illustrate their very different photometric member distributions.
Figure \ref{ovl}(a) shows an X-ray system (ID from \citealp{George11} : 29) at $z = 0.34$.
This X-ray systems is an unambiguous isolated system with no other X-ray system within 3 arcmin.
Filled circles in Figure \ref{ovl}(a) indicate the 119 photometrically identified members
 and open circles are galaxies with $|\Delta c(z_{galaxy} - z_{system}) / (1 + z_{system})| < 1000~\kms$.
Figure \ref{ovl}(b) displays the R-v diagram of this system;
 red filled circles are the member candidates with $P_{mem} > 0$ and
 black open circles are the galaxies with spectroscopic redshifts.
We identify 14 spectroscopic members of this system and calculate the velocity dispersion of the system.
This X-ray system lies on the X-ray luminosity - velocity dispersion scaling relation of the massive clusters
 (Section \ref{relation}).

Figure \ref{ovl}(c) and (d) demonstrate
 an ambiguous case where several X-ray systems overlap.
The complexity of this region was highlighted in \citet{Knobel09}.
Figure \ref{ovl}(c) shows the spatial distribution of galaxies around four X-ray systems (ID from \citealp{George11} : 191, 193, 201, 321).
These four X-ray systems are essentially at the same redshift $z = 0.22$.
We display the photometrically identified members ($P_{mem} > 0$) color coded by membership in the individual clusters.
Although the membership determination algorithm assigns membership within the individual X-ray systems,
 the spatial distributions of the photometrically identified members overlap significantly on the sky.

In Figure \ref{ovl},
 we show the R-v diagram for the photometrically identified members (filled circles) of these four X-ray systems.
We have spectroscopic redshifts for 64 galaxies among the 154 member candidates.
Among these 64 galaxies,
 40 are within $|\Delta c(z_{galaxy} - z_{system}) / (1 + z_{system})| < 1000~\kms$, where $z_{system} = 0.22$.
These spectroscopically identified members are all well within the $R_{200, X}$s of the X-ray systems.

Figure \ref{ovl}(d) plots the R-v diagram of galaxies in the four overlapping systems based on the X-ray center of the system ID 193.
Because the redshifts of the four X-ray systems are essentially identical,
 separation of the members in the individual clusters is ambiguous even with spectroscopic redshifts.
This plot suggests that
 there may be a single cluster at $z \sim 0.22$ and with four separate X-ray peaks.
In this field, there is a weak excess of galaxies at $z \sim 0.35$ and $\sim 0.74$,
 but these galaxies do not show a strong overdensity on the sky.
Thus, we cannot tell whether the X-ray emission has some optical counterparts at greater redshift.
We note that we do not use this sample for studying $L_{X} - \sigma_{v}$ scaling relation
 because this system is removed by the X-ray flags (see Section \ref{refined}).

\subsection{Membership Calibration}\label{calibration}

We test the membership probability from the COSMOS galaxy catalog with the spectroscopic data.
We first define the spectroscopic member fraction ($f_{Spec. mem.}$):
\begin{equation}
f_{Spec. mem.} = \frac{N_{Spec. mem. \cap P_{mem}}}{N_{Spec. \cap P_{mem}}},
\end{equation}
 where  $N_{Spec. \cap P_{mem}}$ is the number of member candidates with spectroscopic redshifts and
 $N_{Spec. mem. \cap P_{mem}}$ is the number of spectroscopic members.
This fraction measures the number of spectroscopic members among the X-ray system member candidates
 with spectroscopic data.

Overall, regardless of the magnitude,
 $\sim 62\%$ (1144) of the member candidates are identified as spectroscopic members.
We also note that there are 232 spectroscopic members without a membership probability;
 nonetheless they are probable members.

\begin{figure}
\centering
\includegraphics[scale=0.48]{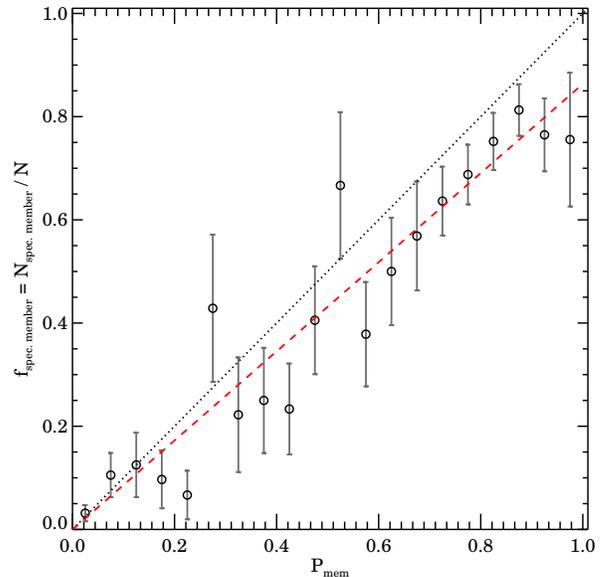}
\caption{Fraction of spectroscopically identified members among the X-ray system members with spectroscopic redshifts as a function of X-ray system membership probability.
Error bars are Poisson.
The dotted line is the one-to-one relation.
The dashed line shows the linear fit. }
\label{pmem}
\end{figure}

Figure \ref{pmem} displays the spectroscopic member fraction as a function of membership probability.
The spectroscopic member fraction generally correlates with the membership probability;
 the spectroscopic member fraction is higher at higher membership probability.
Interestingly, Figure \ref{pmem} shows that
 about 20\% of galaxies with $P_{mem} > 0.9$ are non-members.
Over a wider membership probability range $0.2 < P_{mem} < 1$,
 the number of spectroscopic member is lower than the number expected from the membership probability.

Based on the spectroscopic membership fraction,
 we empirically calibrate the photometric membership probability.
The simple linear fit to the spectroscopic membership fraction yields
\begin{equation}
f_{Spec, mem} = (0.00 \pm 0.01) + (0.88 \pm 0.04) P_{mem}.
\end{equation}

\citet{Sohn18b} examine the spectroscopic membership fraction in the redMaPPer clusters based on the essentially same technique.
They used 104 redMaPPer clusters in HectoMAP redshift survey.
The redMaPPer membership probability also shows a one-to-one relation with the spectroscopic membership fraction,
 but the slope is slightly shallower ($0.66 \pm 0.02$) than for these COSMOS X-ray systems.

\subsection{Total Catalog of COSMOS X-ray Systems}\label{catalog}

Our final sample of COSMOS X-ray systems includes 173 systems with at least one spectroscopic member.
We identify 137 spectroscopic groups with three or more spectroscopic members.
These spectroscopic groups span the redshift range $0.07 < z < 0.99$ and
 the X-ray luminosity range $41.3 < \log L_{X} < 43.9$.

Table \ref{group} compares the COSMOS X-ray system sample and the other recent X-ray selected system catalogs.
AEGIS yields an X-ray group catalog covering similar redshift and X-ray luminosity ranges over a $\sim 0.67$ deg$^{2}$ field of view \citep{Erfanianfar13}.
The typical number of spectroscopic members of these AEGIS groups is six, comparable with COSMOS X-ray systems.
There are 49 AEGIS groups, three times smaller than the number of COSMOS X-ray systems
approximately as expected given the relative area covered by the two surveys.

We also compare the XMM-XXL survey \citep{Pacaud16, Adami18} covering two 25 deg$^{2}$ fields.
\citet{Adami18} identify 365 X-ray systems including 260 spectroscopically identified groups with three or more spectroscopic members
 within the redshift range $0.0 < z < 1.2$.
The XXL groups cover an X-ray luminosity range similar to the COSMOS X-ray systems.
Because the XXL survey flux limit is a few times brighter than the COSMOS and AEGIS X-ray surveys,
 the XXL survey yields a lower X-ray system surface number density.

\begin{deluxetable*}{lccccc}
\tablecolumns{6}
\tabletypesize{\scriptsize}
\setlength{\tabcolsep}{0.05in}
\tablecaption{X-ray Systems in Various Surveys}
\tablehead{
\colhead{Survey} & \colhead{FoV} & \colhead{$\log L_{X}$ range} & \colhead{z range} &
\colhead{$N_{system}$\tablenotemark{a}} & \colhead{$N_{spec. group}$\tablenotemark{b}}}
\startdata
COSMOS\tablenotemark{c}				& 1.65	& [41.3, 43.9] & [0.07, 0.99] & 173		& 137 \\
refined COSMOS\tablenotemark{c}	& 1.65	& [41.3, 43.9] & [0.10, 0.70] & \nodata	&   74 \\
AEGIS \tablenotemark{d}					& 0.67	& [41.7, 43.9] & [0.06, 1.55] &   49		&   47 \\
XXL \tablenotemark{e}						& 50.0	& [41.4, 44.4] & [0.00, 1.20] & 365		& 260
\enddata
\label{group}
\tablenotetext{a}{Total number of systems with one or more spectroscopic redshift.}
\tablenotetext{b}{Total number of spectroscopically identified groups with three or more spectroscopic redshift.}
\tablenotetext{c}{This study}
\tablenotetext{d}{\citet{Erfanianfar13}}
\tablenotetext{e}{\citet{Adami18}}
\end{deluxetable*}

Table \ref{memcat} lists the COSMOS X-ray group members with spectroscopic redshifts in the 173 systems.
We include 1611 X-ray system member candidates with $P_{mem} > 0$ and with a spectroscopic redshift.
We also list the 232 spectroscopic members without $P_{mem}$.
The total number of spectroscopic member is 1843.
This catalog includes
 the X-ray system ID from \citet{George11}, R.A., Decl., spectroscopic redshifts,
 $r-$band magnitude, $P_{mem}$, spectroscopic membership flag, BGG flag,
 and the flag for sample we use for deriving $L_{X} - \sigma_{v}$ scaling relation (Section \ref{scaling}).

\begin{deluxetable*}{lcccccccc}
\tablecolumns{9}
\tabletypesize{\footnotesize}
\setlength{\tabcolsep}{0.1in}
\tablecaption{Spectroscopic Catalog of  COSMOS X-ray System Members }
\tablehead{
\colhead{X-ray system ID} & \colhead{R.A.} & \colhead{Decl.} & \colhead{z} &
\colhead{r mag} & \colhead{$P_{mem}$} & \colhead{Spec. Membership} & \colhead{BGG?} & \colhead{$L_{X}-\sigma_{v}$ Sample}}
\startdata
G11\_011 & 150.130496 &   1.622421 &  0.184000 &  21.30 & 0.48 & N & N & Y \\
G11\_011 & 150.121852 &   1.672978 &  0.217700 &  20.99 & 0.84 & Y & N & Y \\
G11\_011 & 150.114914 &   1.668760 &  0.220100 &  21.60 & 0.89 & Y & N & Y \\
G11\_011 & 150.147597 &   1.632191 &  0.227550 &  19.86 & 0.84 & N & N & Y \\
G11\_011 & 150.155999 &   1.712134 &  0.216000 &  20.13 & 0.80 & Y & N & Y \\
\enddata
\label{memcat}
\tablecomments{The entire table is available in machine-readable form in the online journal. }
\end{deluxetable*}

\subsection{Refined COSMOS X-ray Systems for the $L_{X} - \sigma_{v}$ Scaling Relation}\label{refined}

We investigate the $L_{X} - \sigma_{v}$ scaling relation using the COSMOS X-ray systems (Section \ref{scaling}).
In order to investigate the $L_{X} - \sigma_{v}$ scaling relation,
 we construct a cleaner sample based on X-ray flags and the completeness of the redshift measurements.
The total sample of COSMOS X-ray systems includes
 some X-ray systems with uncertain X-ray luminosity
 due to the overlapping X-ray systems or to imaging defects.
Moreover, the spectroscopy is not complete for some systems.

We first select COSMOS X-ray systems with an X-ray quality flag ($Q_{X}$) 1 or 2.
Additionally, we select systems with X-ray {\it FLAG\_MERGER} and {\it FLAG\_MASK} equal to zero.
{\it FLAG\_MERGER = 0} indicates no overlap with other X-ray systems and
{\it FLAG\_MASK = 0} indicates that less than 10\% of area in the optical imaging of the X-ray system
 is affected by image defects including saturated stars or the edge of the imaging.
We also require that the systems have at least three spectroscopic members
 and that the spectroscopic completeness to $r = 21.3$ exceeds 50\%
 to avoid systematic effects due to varying incompleteness of the survey.

There are 74 COSMOS X-ray systems satisfying these criteria (Table \ref{group})
These refined COSMOS X-ray systems span the redshift range $0.1 < z < 0.7$; one exception is at $z = 0.97$.
The refined COSMOS X-ray sample is a factor of five larger than 
 the sample of AEGIS X-ray groups used for deriving the $L_{X} - \sigma_{v}$ scaling relation
 \citep{Erfanianfar13}. 

\begin{figure}
\centering
\includegraphics[scale=0.49]{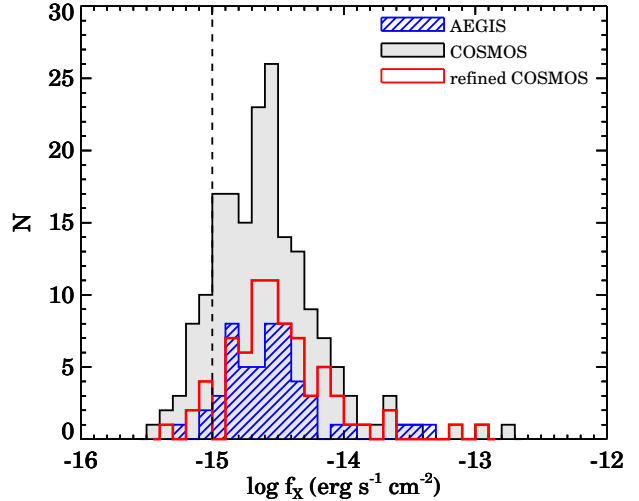}
\caption{
Distribution of X-ray flux for COSMOS X-ray systems (black filled histogram) and
 for the refined COSMOS X-ray systems (red open histogram) we used for studying $L_{X} - \sigma_{v}$ scaling relation.
For comparison, we show the X-ray flux distribution of AEGIS X-ray systems (blue hatched histogram).  }
\label{xflux}
\end{figure}

Figure \ref{xflux} show the X-ray flux distribution of COSMOS X-ray systems and the refined sample.
For comparison, we show the X-ray flux distribution of all of the AEGIS X-ray systems \citep{Erfanianfar13}.
Most of systems in the refined sample have an X-ray flux exceeding
 the COSMOS X-ray survey flux limit ($4\sigma$ detection at $1.0 \times 10^{-15} \ergcms$).
COSMOS X-ray systems with X-ray fluxes near the limit are excluded by the selection process.
The refined COSMOS X-ray systems have an X-ray flux distribution similar to that for AEGIS X-ray systems.

\subsection{Brightest Group Galaxies}\label{bggidentification}

Based on the spectroscopy,
 we identify the brightest galaxy in each X-ray system.
Hereafter, we refer to these brightest galaxies as brightest group galaxies (BGGs).
Identification of the BGGs is important for investigating the cluster dynamics.
The BGG often sits at the bottom of a local potential well (e.g. \citealp{Beers83, Newman13}).

Previous studies of COSMOS X-ray systems define the BGG
 as the galaxy with the largest stellar mass among member candidates within $R_{200}$ of the X-ray center
 \citep{George11, Gozaliasl19}.
We refer to these galaxies with the largest stellar mass as the most massive galaxies, hereafter.
However, in some cases, the most massive galaxy is not the brightest galaxy in the system in the $r-$band.
Furthermore, a large fraction ($\sim 30\%$) of the most massive galaxies are member candidates identified
 based only on photometric redshifts \citep{Gozaliasl19}.

We identify the BGGs of 160 COSMOS X-ray systems
 based on their $r-$band luminosity following the conventional definition of the brightest cluster (group) galaxy
 (e.g. \citealp{Lauer14}).
These 160 COSMOS X-ray systems all contain at least one spectroscopic members within $R_{proj} < 0.5 R_{200}$,
 where $R_{proj}$ is the projected distance from the X-ray peak \citep{Gozaliasl19}.
When we test the BCG identification within $R_{proj} < R_{200}$,
 even brighter galaxies located at large projected distance $0.5 < R_{proj}/R_{200} < 1.0$ are identified as BGGs for 40 systems.
The typical magnitude difference between the BGG and brighter galaxies in the outskirts is $\sim 0.4$ mag.
However, the connection between these bright galaxies in the system outskirts and the X-ray emission is unclear.
We use the tight projected radius $R_{proj} < 0.5 R_{200}$ for BGG identification.

We compare these BGGs with the most massive galaxies identified based on our stellar mass estimates.
There are 30 X-ray systems where the BGG is apparently not the most massive galaxy among the spectroscopic members.
We lack a stellar mass estimate for the BGG of 3 X-ray systems (ID: 11, 196, 264).
\citet{George11} identified these BGGs as the most massive galaxies
 based on their stellar mass estimates with photometric redshifts \citep{George11}.
In 17 systems,
 the BGGs are located closer to the X-ray center than the most massive galaxies,
 suggesting that  the BGGs we identify are closer to the bottom of the potential well.
For the remaining 10 systems,
 the most massive galaxies are closer to the X-ray center than the BGGs.
Large uncertainty in the stellar mass estimate may contribute to the confusion.
In total, the BGGs are also the most massive galaxies in 130 X-ray systems ($\sim 81\%$ of a total 160 COSMOS X-ray systems).

\begin{figure}
\centering
\includegraphics[scale=0.49]{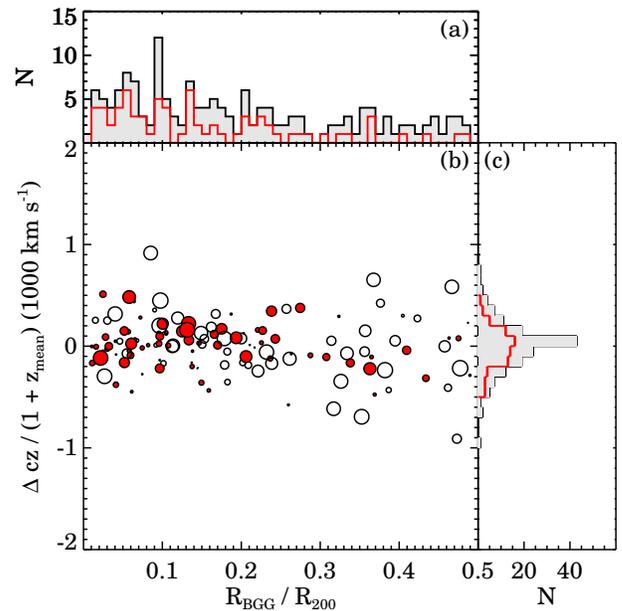}
\caption{
(a) Projected distance distribution between the BGG and the X-ray peak of COSMOS systems.
(b) Relative location of the BGG with respect to the X-ray peak in the R-v diagram.
We take the mean redshift of the member candidates as the central redshift of the X-ray peak.
Black circles indicate the BGG in each of the COSMOS X-ray systems.
Red filled circles highlight the BGG of the COSMOS X-ray systems we use for studying $L_{X} - \sigma_{v}$ scaling relation in Section \ref{scaling}.
Symbol sizes represent the X-ray luminosity of the host X-ray systems; larger symbol indicates brighter X-ray emission.
(c) The distribution of relative radial velocity differences between the BGGs and the central redshift. }
\label{bgg}
\end{figure}

We next examine the BGG offset from the X-ray center.
\citet{Gozaliasl19} studied the offset between the most massive galaxies and the X-ray center of the COSMOS X-ray systems.
They investigated several relations between the BGG offset and other group properties
 including group redshift, X-ray flux, and the magnitude gap between the first and the second brightest galaxies.
Here, we limit our investigation to the offset between the X-ray centers and the BGGs identified based on the luminosity.
We use this indicator as a verification for the refined sample of X-ray systems.

Figure \ref{bgg}(a) shows the distribution of projected distances between BGGs and the X-ray center
 normalized by $R_{200}$ of the individual system;
 the red histogram shows the distribution for the 74 refined COSMOS X-ray systems.
For the refined catalog, their BGGs are more concentrated toward the X-ray center.

The BGGs show a larger offset from the X-ray center  than brightest cluster galaxies (BCGs) in massive systems.
\citet{Lavoie16} examine the BCG offset from the X-ray centers based on XXL clusters with a typical mass of $10^{14} M_{\odot}$,
 a factor of four larger than the typical mass of COSMOS X-ray systems.
They show that $\sim 80\%$ of the BCGs in the XXL clusters lie within $R_{proj} < 0.05 R_{500} \sim 0.035 R_{200}$ of the X-ray centers.
However, only 21\% of BGGs in COSMOS X-ray systems are located within $R_{proj} < 0.035 R_{200}$.

Figure \ref{bgg}(b) displays the location of the BGGs with respect to the X-ray center in the R-v diagram.
Red filled circles highlight the BGGs of the refined COSMOS X-ray systems.
Here, we use the group mean redshift as the redshift of the X-ray system.
The sizes of the symbols indicate the X-ray luminosity of the systems;
 a bigger symbol denotes a more luminous X-ray system.
The BGGs of luminous X-ray systems are sometimes in the outskirts of the systems
 as are the BGGs of some low X-ray luminosity systems.

The BGGs generally have a small line-of-sight velocity offset relative to the cluster mean (Figue \ref{bgg} (c)).
There are only 32 systems ($\sim 20\%$) 
 where the rest-frame radial velocity difference between BGG and the cluster mean is larger than $300~\kms$
 (e.g. the typical velocity dispersion of galaxy groups).
The mean redshift of some groups are not well determined due to poor sampling.
Additionally, some systems show evidence of substructures in the R-v diagram;
 these substructures can cause deviation from the mean system redshift
 especially when the objects are sparsely and/or non-uniformly sampled.

The offsets of the BGG may have an important impact on velocity dispersion measurement for the systems (e.g. \citealp{Skibba11}).
If the BGGs are offset from the bottom of the gravitational potential well,
 the one-dimensional velocity dispersion measurement of the system can be biased.
This biased velocity dispersion measurement then affects
 the study of the $L_{X} - \sigma_{v}$ scaling relation (Section \ref{scaling}).
The refined COSMOS X-ray systems  decreases the impact of this issue: $\sim 86\%$ of
 the BGGs have a redshift within $< 300~\kms$ of the system mean (see also Section \ref{challenge:vdisp}). 

\section{THE $L_{X}- \sigma_{v}$ SCALING RELATION}\label{scaling}
Simple theoretical calculations suggest $L_{X} \propto \sigma_{v}^{4}$ \citep{Solinger72, Quintana82},
 where $L_{X}$ is the X-ray luminosity and the $\sigma_{v}$ is the velocity dispersion of a system.
Many observational studies derive a slope ($\alpha$) of the scaling relations for rich clusters
 consistent with theoretical expectations \citep{OrtizGil04, Popesso05, Zhang11, Rines13, Clerc16}.
For $\log L_{X} \gtrsim 43$, the scatter in the scaling relation
 results from variation in the detailed dynamical state of the systems \citep{Zhang11}
 and/or from sparse sampling of the member galaxies.

In contrast with the scaling relation for massive systems,
 the slope of the scaling relations for low X-ray luminosity groups ($41.4 < \log L_{X} < 42.5$) remains uncertain
 (see Section \ref{challenge:ambiguity}).
Several studies report totally different slopes for low X-ray luminosity systems;
 e.g. from $\alpha \sim 4.3$ from \citet{Mulchaey98} to $\alpha \sim 0.4$ from \citet{Mahdavi00}.

The COSMOS X-ray systems provide a unique opportunity for extending the X-ray scaling relation to very low X-ray luminosity
 ($41.4 < \log L_{X} <42.5$).
The large spectroscopic dataset enables measurement of the velocity dispersions of an essentially complete set of systems.
We show the $L_{X} - \sigma_{v}$ scaling relation of COSMOS X-ray systems in Section \ref{relation}
 and discuss the redshift evolution of the scaling relation in Section \ref{evol}.

\subsection{The $L_{X}- \sigma_{v}$ Scaling Relation of COSMOS X-ray Systems}\label{relation}

We investigate the $L_{X} - \sigma_{v}$ scaling relation
 based on the  refined COSMOS X-ray systems (Section \ref{refined}) covering the redshift range $0.1 < z < 0.7$.
For these 74 X-ray systems,
 we  calculate the rest-frame radial velocity differences between the spectroscopic members within $R_{200}$
 and the BGGs of the individual clusters.
We use the bi-weight recipe \citep{Beers90} for computing the velocity dispersions
 following \citet{Popesso07}, a comparison sample.
We discuss the impact of the velocity dispersion determination in deriving the scaling relation in Section \ref{challenge:vdisp}.
We note that according to previous studies the  $L_{X} - \sigma_{v}$ scaling relation in this redshift range shows little or no evolution
 (e.g. \citealp{Girardi01}, see Section \ref{evol}).

\begin{figure}
\centering
\includegraphics[scale=0.48]{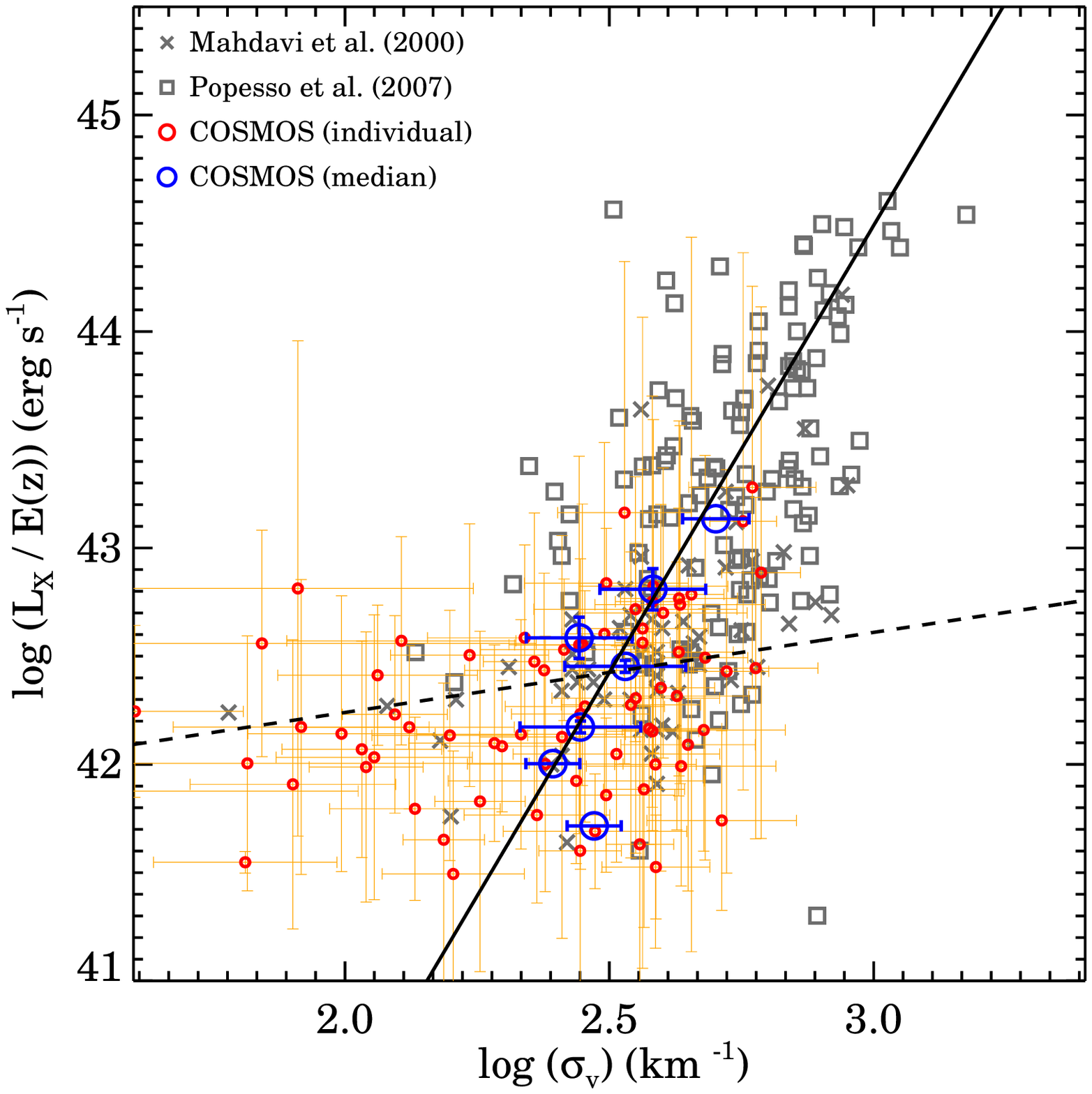}
\caption{$L_{X}$ vs. $\sigma_{v}$ relation.
Gray squares are 137 Abell clusters from \citet{Popesso07} and
Gray crosses are 59 X-ray systems from \citet{Mahdavi00}.
The solid line is the scaling relation derived by \citet{Popesso05}.
The dashed line is the shallower scaling relation for low velocity dispersion systems derived by \citet{Mahdavi00}.
Red circles are the individual COSMOS X-ray systems and
 blue larger circles are the median velocity dispersion within X-ray luminosity bins. }
\label{lxsig}
\end{figure}

Figure \ref{lxsig} displays the $L_{X}-\sigma_{v}$ relation for individual COSMOS X-ray systems (red circles).
Gray squares show the 137 Abell clusters from \citet{Popesso07}
 with X-ray luminosities measured within the energy band $0.1-2.4$ keV and
 galaxy velocity dispersions estimated based on SDSS spectroscopic data.
Gray crosses display the 59 low X-ray systems from \citet{Mahdavi00};
 the X-ray luminosites of these systems are measured within the same energy band.
The solid line is the scaling relation derived from \citet{Popesso05}.
The COSMOS X-ray systems overlap both the sample of \citet{Popesso05} and \citet{Mahdavi00}.
The scatter for the COSMOS system scaling relation is very large for $\log L_{X} < 42.5$
 with an extended tail toward low velocity dispersion.
Several previous studies show similar behavior
 including the extension to low velocity dispersion for low X-ray luminosity systems \citep{Xue00, Osmond04}

Measuring the scaling relation for the X-ray luminosity range $41.5 < \log L_{X} < 42.5$ is challenging.
\citet{Mahdavi00} suggest that X-ray systems in this X-ray luminosity range with $\sigma_{v} < 340~\kms$
 fit a shallow scaling relation $L_{X} \propto \sigma_{v}^{0.37}$
 (dashed lines in Figure \ref{lxsig}).
At first glance, the COSMOS X-ray systems in this X-ray luminosity range also appear to fit this shallow scaling relation.
However, the shallow scaling relation depends on a small number of systems with $\sigma_{v} \lesssim 125~\kms$:
 two systems from \citet{Mahdavi00} and 14 systems from the COSMOS X-ray sample.
The systems with $\sigma_{cl} > 125~\kms$ are scattered around the standard steeper scaling relation that fits more massive systems.

We next compute the more robust median velocity dispersion of the COSMOS X-ray systems
 as a function of their X-ray luminosity.
We divide the set of 74 COSMOS X-ray systems into seven bins according to X-ray luminosity.
Each bin includes 10 X-ray systems except the lowest X-ray luminosity bin that includes 14 systems.
The median velocity dispersion of the COSMOS X-ray systems (large blue circles in Figure \ref{lxsig})
 is consistent with the velocity dispersion expected from the extrapolation of the scaling relation
 for the high X-ray luminosity systems at a given X-ray luminosity; 
 the overall slope of the relation (fitted to the median dispersion from \citealp{Popesso07} and COSMOS data) is $\alpha = 4.7 \pm 0.7$
One consistent interpretation of the data is that they are all consistent with the expected $L_{X} \propto \sigma_{v}^{4}$.

\subsection{Redshift Evolution of $L_{X} - \sigma_{v}$ Relation}\label{evol}

The COSMOS X-ray systems we use for studying the scaling relation
 are distributed over the redshift range  $0.1 < z < 0.7$.
Galaxy systems evolve significantly over this wide redshift range.
For example,
 N-body simulations show that galaxy systems increase their mass by a factor of $2-3$ (e.g. \citealp{Fakhouri10}).
This mass growth increases the velocity dispersion.
On simple theoretical grounds, the X-ray luminosity and temperature are physically related to the velocity dispersion.
Thus, both the X-ray luminosity and the velocity dispersion should continue to reflect
 the depth of the underlying potential well \citep{Bryan98} and
 the predicted relation between the two measurements should continue to hold.

Previous studies show that indeed there is no significant evolution in the $L_{X} - \sigma_{v}$ scaling relations for galaxy clusters.
\citet{Borgani99} measured the velocity dispersion of 16 galaxy clusters at $0.17 < z < 0.55$ and showed that
 the $L_{X} - \sigma_{v}$ scaling relation of these systems is consistent with that for local clusters.
Similarly, \citet{Girardi01} compared the
 the $L_{X} - \sigma_{v}$  and $T_{X} - \sigma_{v}$ scaling relations
 of local clusters at $z < 0.15$ and higher redshift clusters within the range $0.15 < z < 0.9$ and a mean redshift 0.3.
They found no evidence of redshift evolution of these scaling relations.
\citet{Ruel14} show that there is no evolution in the
 $T_{X} - \sigma_{v}$ relation for $\sim 30$ South Pole Telescope (SPT) clusters at $z < 1.3$.

We also find no evidence of redshift evolution in the $L_{X} - \sigma_{v}$ scaling relation for the COSMOS X-ray systems.
Following previous studies,
 we divide the COSMOS systems into two redshift bins ($0.1 < z < 0.4$ and $0.4 < z < 0.7$) and
 compare the $L_{X}- \sigma_{v}$ relations.
The X-ray systems in the two redshift bins show no significant difference in the scaling relation
 but the scatter is large.
This result is insensitive to the redshift limit where divide the sample.

We also compare the distribution of X-ray systems within narrow redshift bins ($\Delta z = 0.1$).
The higher redshift X-ray systems appear at higher X-ray luminosity
 because the sample is X-ray flux limited.
The X-ray systems in each redshift bin show
 a broad velocity dispersion distribution within a narrow X-ray luminosity range.
In effect, the scaling relation derived based on the median velocity dispersion (blue squares in Figure \ref{lxsig})
 is sorted by redshift;
 the higher X-ray luminosity sample is a higher redshift sample.
Thus, the consistency in the scaling relations of the COSMOS X-ray systems in various redshift ranges and the local systems
 indicates that the scaling between X-ray luminosity and velocity dispersion
 reflects the underlying cluster physics in a way that is independent of redshift in this range.

\section{DISCUSSION}\label{discussion}

We explore the $L_{X} - \sigma_{v}$ scaling relation
 based on COSMOS X-ray systems with a clean X-ray luminosity measurement and with complete spectroscopic coverage.
Although we use well-defined X-ray systems,
 measuring a robust scaling relation is challenging
 because of large uncertainties in the measurements particularly for low X-ray luminosity systems.
Here, we discuss challenges in the estimation of the velocity dispersion of the systems (Section \ref{challenge:vdisp})
 and other ambiguities in deriving the $L_{X} - \sigma_{v}$ scaling relation
 for systems with $41.4 < \log L_{X} < 42.5$ (Section \ref{challenge:ambiguity}).

\subsection{Challenges in Deriving $\sigma_{v}$}\label{challenge:vdisp}

In Figure \ref{lxsig},
 we show the $L_{X} - \sigma_{v}$ scaling relation
 based on the unique sample of 74 X-ray luminosity systems in the COSMOS field within $41.4 < \log L_{X} < 43.3$.
The median velocity dispersion for all of the COSMOS X-ray systems fits the relation derived from luminous X-ray clusters.
However, systems with X-ray luminosity of $\log L_{X} \sim 42$
 have velocity dispersion within the wide range 50 to $500~\kms$.

Here, we explore several issues that affect the velocity dispersion of the systems
 and discuss their impact on the $L_{X} - \sigma_{v}$ scaling relation.
We first investigate the impact of velocity dispersion measurement recipes.
We use the velocity dispersion measured based on the  bi-weight recipe \citep{Beers90}
 following measurements made previously.
This recipe works well for estimating the velocity dispersion of systems with small numbers of members.
However, this recipe does not account for uncertainties in the individual redshift measurements.
We thus compare the bi-weight velocity dispersion measurements with measurements
 based on the recipe from \citet{Danese80} that takes measurement error into account.
The velocity dispersion measurements from these two recipes are consistent with one another,
 indicating that the velocity dispersion measurements are robust.

\begin{figure*}
\centering
\includegraphics[scale=0.50]{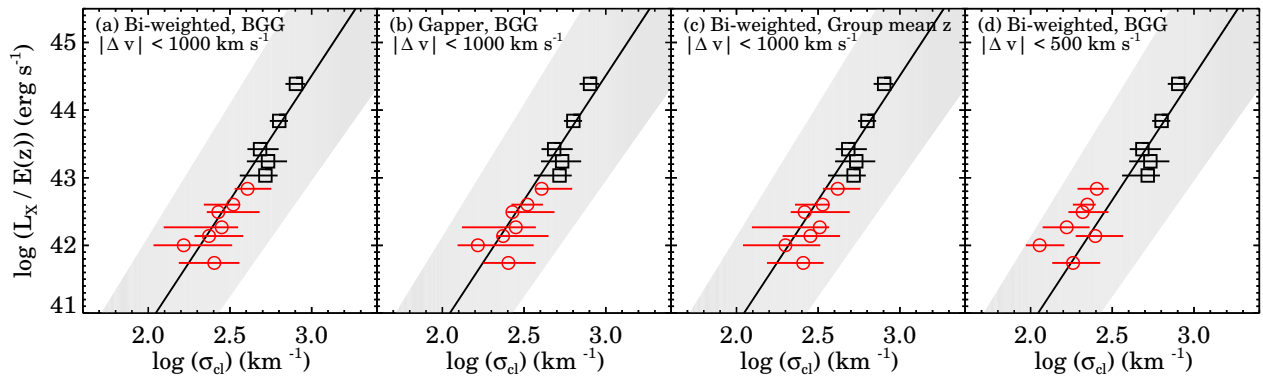}
\caption{Test of $L_{X} - \sigma_{v}$ scaling relation based on various velocity dispersion measures.
Solid lines and gray shaded regions in each panel display the $L_{X} - \sigma_{v}$ scaling relation derived in \citet{Popesso05} and their 95\% confidence area.
Black squares are the median velocity dispersion for the Abell cluster sample from \citet{Popesso07}. 
(a) Red circles show the median velocity dispersion for the refined COSMOS X-ray sample
 based on the bi-weight method.
The error bars show the interquartile range. 
Bins in X-ray luminosity are as in Figure \ref{lxsig}.
The dashed line is a fit to the Abell cluster and the COSMOS X-ray systems: $L_{X} \propto \sigma_{v}^{4.8}$. 
(b) Same as panel (a), but for velocity dispersion measured with the gapper method.
(c) Same as panel (a), but for velocity dispersion measured centered on the group mean redshift.
(d) Same as penal (a), but for velocity dispersion measured for spectroscopic members within
 the radial velocity limit $|\Delta cz| / (1+z_{BGG}| < 500~\kms$.  }
\label{lxsig_test}
\end{figure*}

We also calculate the velocity dispersion using the ``gapper" estimator method \citep{Beers90},
 another good velocity dispersion estimator for systems with small numbers of members
 \citep{Erfanianfar13, Clerc16}.
Figure \ref{lxsig_test} (a) and (b) compare the $L_{X} - \sigma_{v}$ relation
 based on velocity dispersion measurement using the bi-weight and the gapper methods.
The black solid line and the gray shaded area show the $L_{X} - \sigma_{v}$ scaling relation for massive clusters
 and 95\% confidence region \citep{Popesso05}.
Red circles in Figure \ref{lxsig_test} (a) and (b) show the median velocity dispersions for the refined COSMOS X-ray sample bins as in Figure \ref{lxsig}.
The error bars show interquartile range for the velocity dispersion measurements.
The gapper velocity dispersions are systematically larger than the bi-weight velocity dispersions only by $\sim 10\%$.
This systematic difference is much smaller than the uncertainties in individual velocity dispersion measurements.

We also compare velocity dispersion measurements based on different centers.
We assume that the BGGs are located at the center of the gravitational potential of the system.
To test the impact of possible center mis-identification,
 we compute the velocity dispersions centered on the most massive galaxy and
 on the mean redshift of cluster members.

Comparison between the velocity dispersion measurements
 based on the BGG and the mean redshift of group members
 shows no systematic difference.
The $1\sigma$ standard deviation of the ratio between these two velocity dispersion measurements is $\sim 0.24$.
Figure \ref{lxsig_test} (c) shows the $L_{X}-\sigma_{v}$ scaling relation
 based on the velocity dispersion measurement centered on the mean redshift of group members.

We finally examine the impact of the radial velocity limit for identifying spectroscopic members.
The radial velocity limit separates cluster members and line-of-sight interlopers.
If the radial velocity limit is too large,
 interlopers with large relative radial velocity can be identified as members
 and inflate the velocity dispersion, and vice versa.
The caustic method \citep{Diaferio97, Diaferio99, Serra13} is often used for identifying cluster members
 to minimize the contamination by interlopers.
However, we cannot apply the caustic technique because the typical number of spectroscopic members in COSMOS X-ray systems is too small.

We test sensitivity of the velocity dispersions of the COSMOS X-ray systems
 by changing the radial velocity limit ($|\Delta c(z - z_{sys}) / (1+z_{sys})|$) from 500 to $2000~\kms$.
Figure \ref{lxsig_test} (d) displays the $L_{X} - \sigma_{v}$ scaling relation
 based on velocity dispersion measurements with a tight radial velocity limit of $500~\kms$.
The velocity dispersion of systems with $\sigma_{v} > 300~\kms$ are reduced significantly.
The tight radial velocity limit only affects systems with high velocity dispersion
 by excluding possible member candidates with large radial velocity difference.
The velocity dispersion measurements changes little with generous radial velocity limits $> 700~\kms$.

Interestingly, the entire $L_{X} - \sigma_{v}$ scaling relation derived based on the massive clusters and the less massive COSMOS sample
 is insensitive to the radial velocity limits we use.
The variation in radial velocity limit affects only the large velocity dispersion systems.
However, the $L_{X} - \sigma_{v}$ scaling relation of COSMOS X-ray systems
 based on the reduced velocity dispersions with a tight radial velocity limit
 still lie on the scaling relation defined by massive clusters.
For the low velocity dispersion systems ($\sigma_{v} < 300~\kms$),
 the velocity dispersion measurement is insensitive to the radial velocity limits.
\citet{Erfanianfar13} show that the same effects for the smaller but similarly defined AEGIS sample.

In summary,
 we test the $L_{X} - \sigma_{v}$ scaling relation for COSMOS X-ray systems
 based on velocity dispersions measured with various approaches.
These variations  affect individual systems
 and contribute to the scatter in the scaling relation.
However, the global $L_{X} - \sigma_{v}$ scaling relation remains within the 95\% confidence range determined from other samples.

\begin{figure}
\centering
\includegraphics[scale=0.48]{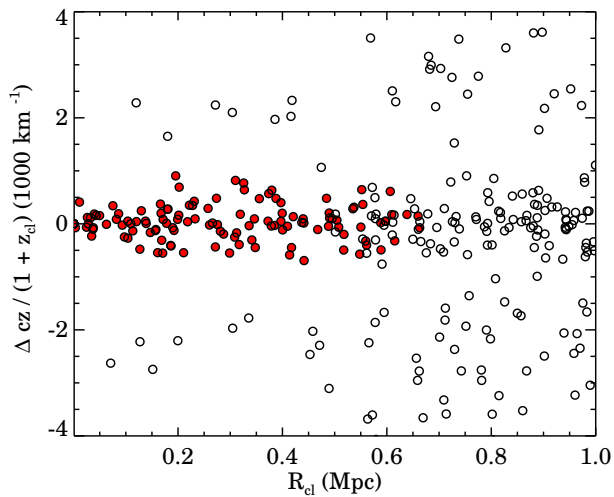}
\caption{Stacked R-v diagram of the 15 lowest X-ray luminosity systems in the COSMOS sample.
Red filled circles are spectroscopic members and the open circles are galaxies with spectroscopic redshifts.
There is no clear difference between the velocity dispersion of X-ray system members and the radial extent of the surrounding large scale structure. }
\label{rvstack}
\end{figure}

Measuring the velocity dispersion is fundamentally challenging
 for low velocity dispersion systems.
All of the COSMOS X-ray systems are embedded in the surrounding large scale structure (see Figure \ref{rv}).
For example, we plot a stacked R-v diagram for galaxies around
 21 X-ray systems we used for deriving the $L_{X}-\sigma_{v}$ scaling relation
 within the X-ray luminosity range $42.0 < \log L_{X} < 42.25$ (Figure \ref{rvstack}).
Galaxies with $R_{cl} \gtrsim 0.6$ Mpc (open circles) inhabit the surrounding large scale structure.

In the R-v diagram,
 the radial velocity distribution of the X-ray system members is undistinguished from
 that of nonmember galaxies in the outskirts.
Massive clusters or groups tend to show a larger
 extension in the radial velocity direction (a finger),
 toward the cluster center (e.g. see the R-v diagram of the massive cluster A2029 in \citealp{Sohn19}).
For a typical low X-ray luminosity system (Figure \ref{rvstack}),
 the width of the radial velocity distribution for  galaxies with $0.6 < R_{proj} < 1.0$ Mpc is $\sim 350~\kms$
 comparable to or exceeding the characteristic velocity dispersion of the stacked cluster members.
Dense sampling of a single system could yield more robust, distinctive measurements for these low velocity dispersion systems.

Measuring the velocity dispersion of systems with intrinsically low velocity dispersion is extremely difficult
 (see more in Section \ref{challenge:ambiguity}).
For these systems, the velocity dispersion cannot be cleanly disentangled from
 the radial velocity dispersion of the surrounding large scale structure.
In effect,
 the effective velocity dispersion of the surrounding structure sets a floor on the velocity dispersion of some poorly sampled X-ray systems.
This issue may contribute to ambiguity in determining of $L_{X} - \sigma_{v}$ slope at low X-ray luminosity.

\subsection{Ambiguity in Determing $L_{X} - \sigma_{v}$ Scaling Relation for Low-Mass Systems}\label{challenge:ambiguity}

The $L_{X} - \sigma_{v}$ relation at low X-ray luminosity presents a puzzle.
There are two very different slopes of the scaling relation for low X-ray luminosity systems in the literature so far:
 1) a similar slope ($\alpha \sim 4 - 5$) to that for high X-ray luminosity clusters versus
 2) a shallower slope ($\alpha \sim 0.4-2$).
Early on,
 \citet{Ponman96} show that the $L_{X} - \sigma_{v}$ relation for 22 Hickson compact groups with $41.0 < \log L_{X} < 43.0$
 has a slope ($\alpha = 4.9$) similar to high X-ray luminosity clusters.
Based on a similar sized sample with similar X-ray luminosity,
 \citet{Mulchaey98} and \citet{Helsdon00} show that
 the scaling relation of poor X-ray groups is consistent with that for galaxy clusters with large scatter.
\citet{Erfanianfar13} also showed that the scaling relation for X-ray groups in the AEGIS survey
 scatter around the scaling relation for X-ray luminous clusters.

In contrast, \citet{Mahdavi00} suggest that a broken power law is
 the best fit to the $L_{X} - \sigma_{v}$ relation of low X-ray luminosity systems
 (see also \citealp{Dell'Antonio94}).
They derived a much shallower scaling relation ($L_{X} \propto \sigma_{v}^{0.37}$)
 for systems with $\sigma_{v} < 340~\kms$.
Other studies based on the larger number of galaxy groups also report a shallower scaling relation
 ($\alpha  = 2.3 - 3.3$, \citealp{Xue00, Osmond04, Brough06, Khosroshahi07}).

The discrepancies result from a small number of low X-ray luminosity systems
 with extremely low velocity dispersion ($\sigma \lesssim 125~\kms$).
The shallow scaling relations reported in previous studies
 are derived based on X-ray samples including these extreme systems.
Indeed, the 14 COSMOS systems ($\sim 16\%$) with very low velocity dispersion
 appear to fit the shallow scaling relation.

The physical origin of these extremely low velocity dispersion system is unclear.
\citet{Mamon99} argued that the minimum velocity dispersion of a virialized galaxy system
 should be larger than the sum of the masses of its member galaxies.
Surprisingly, the velocity dispersions of outliers in the COSMOS X-ray sample are sometimes smaller than
 the stellar velocity dispersion of their BGGs (J.Sohn et al. 2019, in preparation).
\citet{Helsdon05} suggested three possible explanations for reduced velocity dispersions;
(1) dynamical fraction reduces the orbital velocities,
(2) tidal interaction reduces the orbital velocities of the galaxies, and
(3) the line-of-sight velocity dispersion appears reduces as a result of velocity and/or spatial anisotropy.
The low velocity dispersion systems in the COSMOS field may result from one or more of these mechanisms.
Additionally, poor sampling can still result in artificially low velocity dispersions.

\begin{figure}
\centering
\includegraphics[scale=0.47]{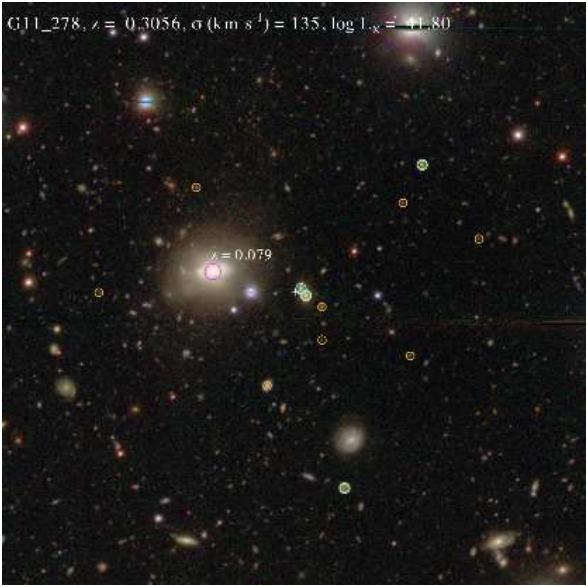}
\caption{HSC image of the X-ray system ID: 278. The image is centered on the X-ray center from \citet{George11}.
Yellow circles are member candidates with $P_{mem} > 0$ and
 cyan circles are spectroscopic members around the central redshift of the system suggested by \citet{George11}.
The magenta circle highlights a foreground galaxy with remarkable shell structure, the likely X-ray source.}
\label{confusion}
\end{figure}

Furthermore, the X-ray source may be misidentified.
ID:278 provides a case in point
 (we note that \citet{Gozaliasl19} exclude this system but it is a kind of prototype for potential problems in the system identification).
Figure \ref{confusion} shows a {\it Subaru}/HSC image of this system.
\citet{George11} identify the apparent red sequence of a galaxy group at $z = 0.306$ associated with X-ray emission.
However, a foreground galaxy with an impressive shell structure ($z = 0.079$) is probably the X-ray source.
In fact the $z =0.306$ apparent system is offset from the center of the X-ray emission
 but the shell galaxy is coincident with it.
We also suggest below that this source is an example of X-ray emission from an individual galaxy.

\begin{figure}
\centering
\includegraphics[scale=0.47]{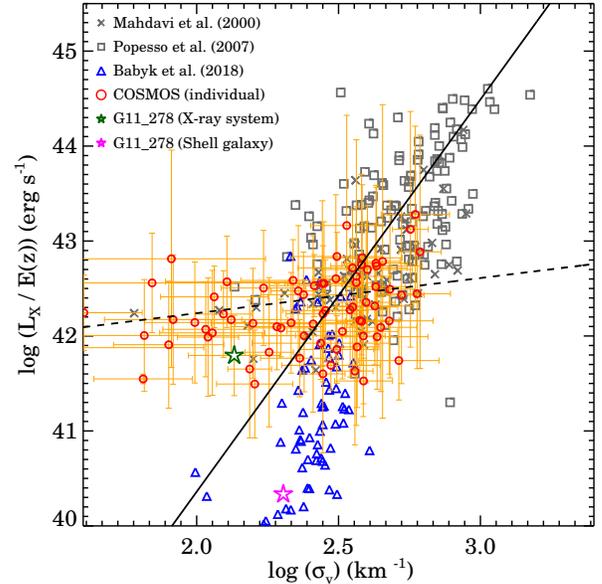}
\caption{$L_{X} - \sigma_{v}$ scaling relation of X-ray systems and early-type galaxies with X-ray emission.
The lines and symbols are the same as in Figure \ref{lxsig}.
Blue triangles are early-type galaxies from \citet{Babyk18};
 the velocity dispersions are the central stellar velocity dispersions within 1 kpc.
The green star symbol is the location of the X-ray system in Figure \ref{confusion} based on the
$z = 0.306$ system velocity dispersion.
The magenta star symbol is the source position if the X-ray emission is associated with the shell galaxy.}
\label{lxsig_galaxy}
\end{figure}

Another ambiguity in deriving the $L_{X} - \sigma_{v}$ scaling relation for low X-ray luminosity system
 is disentanglement of X-ray emission associated with an individual galaxy that may be either
 (1) a system member or (2) a foreground object as in the case of ID:278.
Figure \ref{lxsig_galaxy} compares the $L_{X} - \sigma_{v}$ scaling relation for X-ray systems
 and that for early-type galaxies with X-ray emission.
We use the early-type galaxy sample from \citet{Babyk18}
 that includes {\it Chandra} X-ray luminosities and the central stellar velocity dispersions within 1 kpc.
These early-type galaxies have a substantial overlap with  the X-ray luminosity range of the COSMOS X-ray systems.
At low X-ray luminosity the relation between $L_x$ and $\sigma$ appears to split into to tail.
One associated with poor groups extends to low velocity dispersion with a shallow slope.
The other, corresponding to individual elliptical galaxies is a steep extension.
Understanding this bifurcation is a challenge for the larger X-ray and spectroscopic samples that will be available in the e-ROSITA era.

Returning to ID:278, the green star symbol in Figure \ref{lxsig_galaxy} shows
 the position of the position of the $z = 0.306$ system in the $L_{X} - \sigma_{v}$ relation.
It lies in the low dispersion tail.
If instead, the source is the shell galaxy, the system moves onto the much steeper $L_{X} - \sigma_{v}$ relation for individual galaxies
 based on the central velocity dispersion of the shell galaxy from SDSS DR14 spectroscopy catalog.
This remarkable ambiguity highlights the difficulty in understanding the physical nature of these X-ray sources.

The slope of the $L_{X}-\sigma_{v}$ scaling relation at low X-ray luminosity range ($41.5 < \log L_{X} < 43$) remains uncertain.
The median velocity dispersions of the systems at a given X-ray luminosity
 and the velocity dispersion of the stacked sample strongly indicate that
 the extrapolation of the $L_{X} - \sigma_{v}$ scaling relation for rich clusters fits the low X-ray luminosity systems.
However, detailed examination of the overlap between  individual X-ray systems and early-type galaxies (as in ID:278) suggests a more complex picture. 

Future surveys with e-ROSITA will refine the $L_{X} - \sigma_{v}$ scaling relation
 based on much larger samples of low X-ray luminosity systems.
The high resolution of e-ROSITA will also reduce  uncertainties in X-ray measurements
 including the X-ray center position, critical to identifying optical counterparts.
Furthermore, deep and dense spectroscopic surveys with 
 {\it Subaru}/Prime Focus Spectrograph \citep{Takada14} and 4MOST \citep{Finoguenov19}
 will yield more robust velocity dispersion measurements of low X-ray luminosity systems.

\section{CONCLUSION}\label{conclusion}

COSMOS X-ray systems are a unique flux-limited sample spanning the redshift range $0.06 < z < 0.99$
 and the X-ray luminosity range $41.3 < \log L_{X} < 43.9$.
The COSMOS X-ray system catalog constructed by \citet{George11} we examine here
 includes X-ray system with X-ray fluxes  $\geq 1.0 \times 10^{-15}$ erg cm$^{-2}$ s$^{-1}$,
 an order of magnitude deeper than the future survey with e-ROSITA \citep{Merloni12}.
The rich dataset obtained from multi-wavelength photometric and spectroscopic observations
 enables investigation of the properties of low luminosity X-ray systems.
These COSMOS low X-ray luminosity systems demonstrate the complexity that
should be unraveled by e-ROSITA.

We identify 1843 spectroscopic members in 180 COSMOS X-ray systems.
This large spectroscopic sample combines all available spectroscopic data for  galaxies in the COSMOS field.
There are 137 ($\sim 76\%$) systems that contain at least three spectroscopic members within $R_{200}$.
The typical number of spectroscopic members per cluster is $\sim 5$.
We also explore 5 redMaPPer systems and 13 CAMIRA systems in the COSMOS field.

We identify the brightest group galaxies among the spectroscopic members.
For $\sim 22\%$ of the systems, the BGG is not the galaxy with the largest stellar mass in the systems.
The typical offset between the BGG and the X-ray peak is $R_{proj} \sim 0.14 R_{200}$,
 larger than the typical offset between BCGs and the cluster X-ray peak ($R_{proj} \sim 0.05 R_{200}$).

The $L_{X} - \sigma_{v}$ scaling for a refined subset including 74 COSMOS X-ray systems
 is consistent with relation derived from high X-ray luminosity systems ($L_{X} \propto \sigma_{v}^{4}$)
 but the error and  the scatter are large.
The scaling relation based on robust velocity dispersion measurements from the median velocity dispersion
 of the sample with similar X-ray luminosity
 lies on the extrapolation of the scaling relation for massive clusters.

We explore several effects on the velocity dispersion measurement
 including the measurement recipes, mis-identification of the central redshift of the system, 
 and the radial velocity limits for identifying spectroscopic members.
None of these significantly impact the global slope of the scaling relation.
Variation in the velocity dispersion measurements only increases the random uncertainties in the scaling relation.

We discuss the fundamental difficulty in deriving the $L_{X} - \sigma_{v}$ relation
 at low $L_{X}$ and low $\sigma_{v}$ due to the nature of low X-ray luminosity systems.
At low X-ray luminosity, poor groups and extended emission associated with individual quiescent galaxies overlap. 
The morphology of the distribution of extended X-ray sources in the $L_{X} - \sigma_{v}$ plane appears complex with a tail 
 that includes poor groups and extends toward low velocity dispersion.
In contrast quiescent galaxies populate a locus steeper than the relation for massive clusters \citep{Babyk18}. 

Interpretation of the group velocity dispersion measurement is challenging
 due to the impact of surrounding large scale structure and the poor sampling.
Furthermore identification of X-ray source is difficult and we demonstrate that on the basis of extensive data a source can move from the low dispersion tail to the locus for sources associated with individual galaxies.

Future survey with e-ROSITA will identify a large number of low X-ray luminosity systems from
a large area on sky with better spatial resolution and with temperature measurements \citep{Finoguenov19}.
Low velocity dispersion systems ($\sigma < 150~\kms$) are abundant
 even in the SDSS spectroscopic sample \citep{Sohn16}.
The e-ROSITA survey will measure X-ray properties of these low velocity dispersion systems and
 will facilitate understanding of the morphology of the $L_{X} - \sigma_{v}$ distribution for poor groups and individual galaxies.
Furthermore, deep and complete spectroscopy with PFS survey and 4MOST survey \citep{Finoguenov19}
 will enable a better measurement of system velocity dispersions and the central velocity dispersions of individual galaxies.
Combining these survey data will elucidate the scaling relations for systems and for individual galaxies and the relationship between them.

\acknowledgments
This paper uses data products produced by the OIR Telescope Data Center, supported by the Smithsonian Astrophysical Observatory.
J.S. is supported by the CfA Fellowship.
The Smithsonian Institution supported the research of M.J.G.
H.J.Z. is supported by the Clay Postdoctoral Fellowship.
This research has made use of NASA's Astrophysics Data System Bibliographic Services.
Observations reported here were obtained at the MMT Observatory, a joint facility of the University of Arizona and the Smithsonian Institution.

Funding for SDSS-III has been provided by the Alfred P. Sloan Foundation,
 the Participating Institutions, the National Science Foundation,
 and the U.S. Department of Energy Office of Science.
The SDSS-III web site is http://www.sdss3.org/.
SDSS-III is managed by the Astrophysical Research Consortium for
 the Participating Institutions of the SDSS-III Collaboration including
 the University of Arizona, the Brazilian Participation Group,
 Brookhaven National Laboratory, University of Cambridge,
 Carnegie Mellon University, University of Florida, the French Participation Group,
 the German Participation Group, Harvard University, the Instituto de Astrofisica de Canarias,
 the Michigan State/Notre Dame/ JINA Participation Group, Johns Hopkins University,
 Lawrence Berkeley National Laboratory, Max Planck Institute for Astrophysics,
 Max Planck Institute for Extraterrestrial Physics, New Mexico State University,
 New York University, Ohio State University, Pennsylvania State University,
 University of Portsmouth, Princeton University, the Spanish Participation Group,
 University of Tokyo, University of Utah, Vanderbilt University, University of Virginia,
 University of Washington, and Yale University.

\appendix
\section{Photometrically Identified Clusters in the COSMOS Field}\label{photcl}

For completeness,
 we briefly explore photometrically identified cluster candidates (redMaPPer and CAMIRA systems) in the COSMOS field (see Figure \ref{spatial}) 
 within the redshift range $0.21 < z < 0.58$.
Detailed analyses of these systems including determination of the spectroscopic redshift of the systems and
 the identification of spectroscopic members of the systems are beyond the scope of this study.
Our goal here is simple comparison between these photometrically identified systems and COSMOS X-ray systems.

There are five redMaPPer systems \citep{Rykoff16} in the COSMOS field (Figure \ref{spatial}):
 RM 5652, RM 10096, RM 13718, RM 24235, RM 52747 (IDs from \citealp{Rykoff16}).
These redMaPPer systems were identified based on a red-sequence detection technique applied to SDSS DR8 photometric data.
RM 24235 corresponds to the X-ray system (G11\_262) within $1\arcmin$.
RM 52747 is located in between two X-ray systems (G11\_145 and G11\_160),
 but the redshifts of the X-ray systems differ from the redshift reported by redMaPPer.
The other three systems do not have X-ray counterparts from \citet{George11}
 because they are outside of the area where \citet{George11} identify the X-ray systems. 
With the larger search area, \citet{Gozaliasl19} identify all of the redMaPPer systems except RM 52747. 
In summary, we identify X-ray counterparts for 80\% of the redMaPPer systems,
 slightly lower than the result based on a larger X-ray cluster survey (86\%, \citealp{Sadibekova14}).

\begin{figure}
\centering
\includegraphics[scale=0.47]{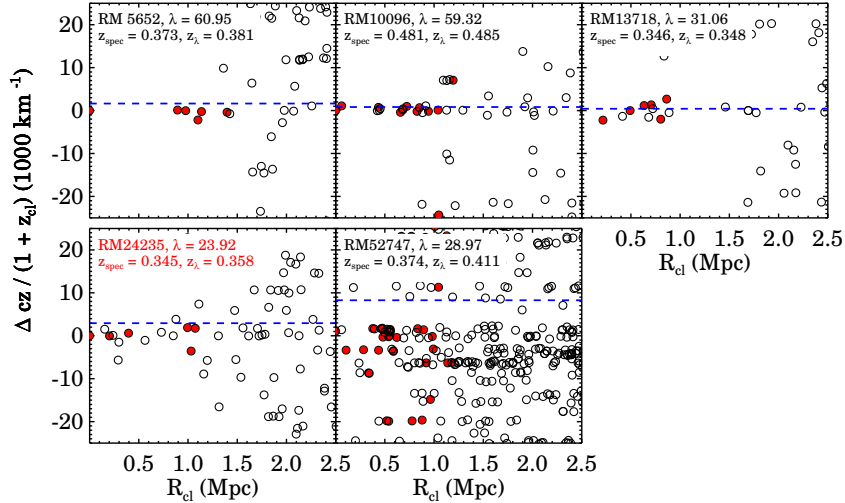}
\caption{R-v diagrams of the five redMaPPer systems in the COSMOS field.
The labels indicate redMaPPer ID, richness ($\lambda$), the median redshift of the redMaPPer member candidates with $P_{mem} > 0.5$,
 and the redshift from the redMaPPer catalog (dashed line).
The system with a red label has an X-ray counterpart from \citet{George11}.
Red filled circles indicate the redMaPPer member candidates with membership probabilities larger than 0.5.
Black open circles are the galaxies with spectroscopic redshifts. }
\label{rv_rm}
\end{figure}

Figure \ref{rv_rm} shows the R-v diagrams of the five redMaPPer systems.
The redMaPPer member galaxy catalog \citep{Rykoff16} provides a photometric membership probability ($P_{mem}$) for each galaxy.
We match the redMaPPer member galaxy catalog and the COSMOS galaxy catalog
 to obtain the redMaPPer membership probability of galaxies in the COSMOS galaxy catalog.
The open circles in Figure \ref{rv_rm} display the galaxies with spectroscopic redshifts
 and the filled circles are the redMaPPer member candidates with $P_{mem} > 0.5$.
We determine the central redshifts of the redMaPPer system
 as the median redshift of the redMaPPer member candidates with $P_{mem} > 0.5$.
The dashed line is the system redshift ($z_{\lambda}$) from the redMaPPer catalog.

There are galaxy overdensities at the redMaPPer centers.
The central redshifts of the redMaPPer clusters are fairly well determined,
 except for one system (RM52747) which is in a complex field.
Similar to the X-ray cluster member candidates,
 several member candidates with high membership probabilities are not spectroscopic members of the systems.
It is interesting that
 there are a number of spectroscopic members without membership probabilities.
These galaxies may be excluded from the redMaPPer membership catalog because they are faint or blue.

The CAMIRA cluster catalog includes 13 candidate systems in the COSMOS field;
 six systems have X-ray counterparts from \citet{George11}.
\citet{Gozaliasl19} identify X-ray counterparts for eight CAMIRA systems 
 in addition to the six systems with X-ray counterparts from \citet{George11}. 
Intriguingly, only three CAMIRA systems match redMaPPer systems
 although both cluster catalogs were based on a red-sequence technique.
The CAMIRA systems without redMaPPer counterparts are mostly (80\%) systems with $z > 0.3$.

\begin{figure}
\centering
\includegraphics[scale=0.47]{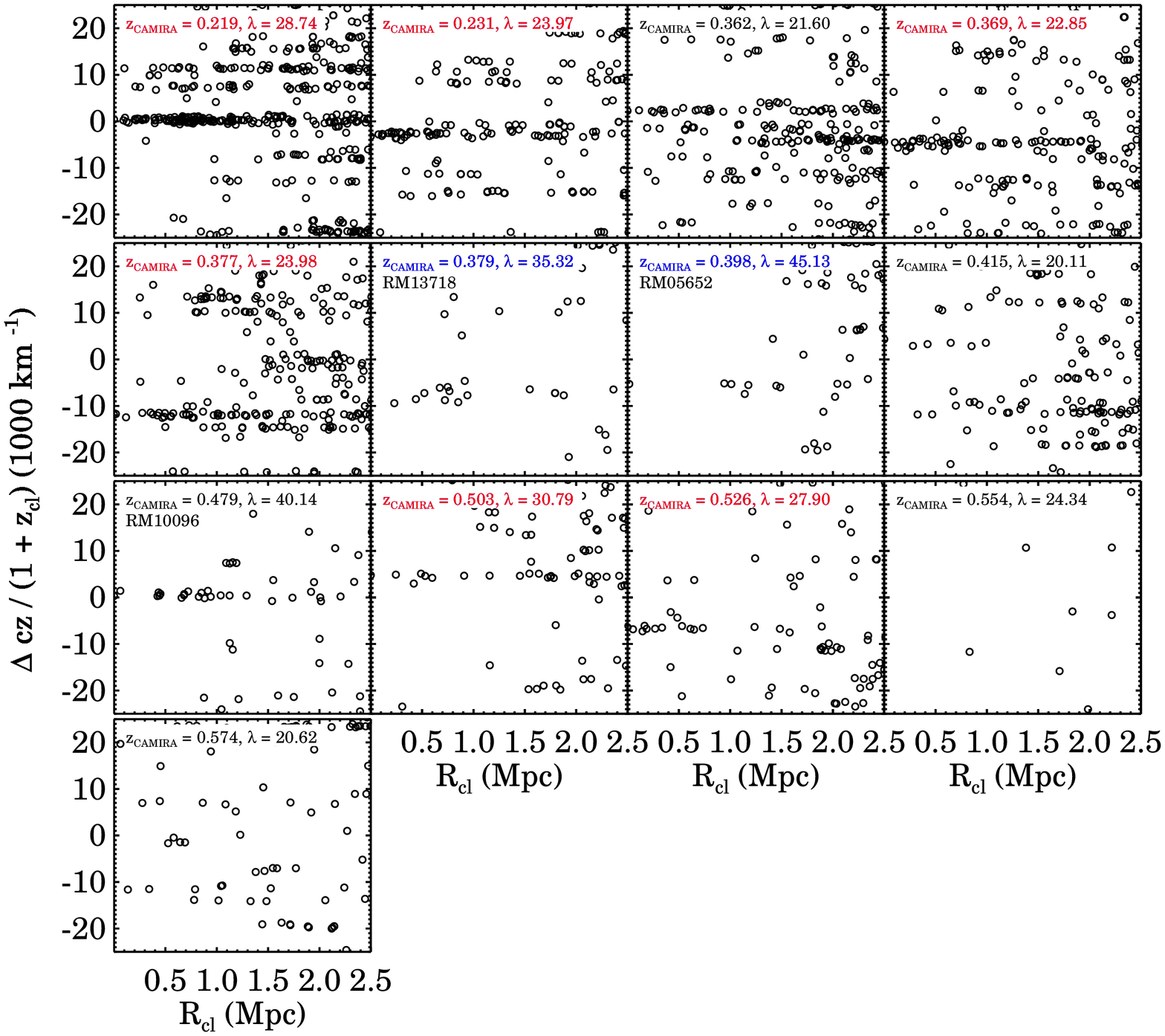}
\caption{R-v diagrams of the 13 CAMIRA candidate systems \citep{Oguri14} in the COSMOS field.
The labels indicate the central redshift and the corrected richness from the CAMIRA catalog.
The systems with red labels have X-ray counterparts from \citet{George11}
 and those with blue labels have X-ray counterparts from \citet{Gozaliasl19}.
The systems with a redMaPPer ID have a redMaPPer counterpart.
Black open circles indicate galaxies with spectroscopic redshifts. }
\label{rv_cm}
\end{figure}

We examine the R-v diagrams of the 13 CAMIRA systems in Figure \ref{rv_cm}.
Unlike the redMaPPer catalog, the CAMIRA catalog does not provide a membership probability.
Thus, we simply use the central redshifts quoted in the CAMIRA catalog to plot the R-v diagram
 instead of the central redshifts determined based on  spectroscopically identified members.
The black open circles indicate  galaxies with spectroscopic redshifts around the CAMIRA cluster center.

There are galaxy overdensities for some CAMIRA clusters,
 but the central redshifts from the catalog are often significantly offset from the overdensities
in redshift space.
We do not see  a galaxy overdensity at the suggested cluster center for 
the systems at $z =0.379, 0.398, 0.554$ and 0.574.
These systems may be false positive or the actual system is at a very different redshift.

Previous studies examine the relation between
 the X-ray luminosity and the richness of optically-identified galaxy clusters:
 \citet{Oguri14} examine CAMIRA systems and \citet{Hollowood18} examine redMaPPer systems.
Based on these relations, we expect that three X-ray systems (ID: 011, 120, 220) have richness larger than 20 suggesting that they should be in the photometric catalogs.
However, these three systems are missing from the redMaPPer catalog and
 only one system (ID: 011) is included in the CAMIRA catalog.
The two missing X-ray systems are perhaps missing from these catalogs because they are at high redshift ($z > 0.72$).
The system ID 011 is a prominent cluster at $z = 0.22$, but the redMaPPer algorithm  misses it.
These differences among the catalogs highlight the challenge of constructing complete catalogs of galaxy systems even for relatively massive systems.

\clearpage

{}

\end{document}